\documentclass[preprintnumbers,superscriptaddress,showkeys,showpacs,byrevtex]{revtex4}
\usepackage{amsmath,amsfonts,amssymb,amscd,amsxtra,amsthm}
\usepackage{graphicx}
\usepackage{xcolor}
\usepackage{epstopdf}
\usepackage{bm}
\preprint{PKNU-NuHaTh-2019-02}
\begin{document}
\title{Studies on photo- and electro-productions of $\Lambda(1405)$ via $\gamma^{(*)} p\to K^{*+}\pi^0\Sigma^0$}
\author{Seung-il Nam}
\email[E-mail: ]{sinam@pknu.ac.kr}
\affiliation{Department of Physics, Pukyong National University (PKNU), Busan 48513, Republic of Korea}
\affiliation{Asia Pacific Center for Theoretical Physics (APCTP), Pohang 37673, 
Republic of Korea}\date{\today}
\author{Atsushi Hosaka}
\email[E-mail: ]{hosaka@rcnp.osaka-u.ac.jp}
\affiliation{Research Center for Nuclear Physics (RCNP), Ibaraki, Osaka 567-0047, Japan}
\affiliation{Advanced Science Research Center, Japan Atomic Energy Agency (JAEA), Tokai 319-1195, Japan}
\date{\today}
\begin{abstract}
We study the photo- and electro-productions of the vector kaon off the proton, i.e.,  $\gamma^{(*)}p\to K^{*+}\pi^0\Sigma^0$, and investigate the line shape of the $\pi^0\Sigma^0$ invariant mass in an effective Lagrangian approach with the inclusion of a $K^*N\Lambda^*$ interaction. Relevant electromagnetic form factors for the neutral hyperons and charged strange mesons are constructed by considering experimental and theoretical information. We find that the $\Lambda^*$ peak is clearly observed for the photo- and electro-productions with the finite $K^*N\Lambda^*$ interaction, whereas the clear peak signals survive only for the electro-production, when we ignore the interaction. These different behaviors can be understood by different $Q^2$ dependences in the $K^*$ electromagnetic and $K^*\to\gamma K$ transition form factors. We suggest a photon-polarization asymmetry $\Sigma$ to extract information of the $K^*N\Lambda^*$ interaction. It turns out that $\Sigma$ near the $\Lambda^*$ peak region becomes negative with a finite $K^*N\Lambda^*$ interaction while positive without it for $Q^2 = 0$, due to the different naturalities of $K$ and $K^*$ exchanges.  For $Q^2\ne 0$, we observe more obvious signals in the peak region due to the additional contribution of the longitudinal virtual photon for $\Lambda^*$. 
\end{abstract}
\maketitle
\section{Introduction}
Hadron spectroscopy is one of the most active fields to understand the nature of the low-energy (nonperturbative) strong interactions of quarks and gluons, being governed by quantum chromodynamics (QCD). In the low-energy region, it is believed that chiral symmetry plays an important role, which provides nontrivial mechanisms to generate the masses of hadrons as known by spontaneous breakdown of chiral symmetry (SBCS). Based on this idea, effective field theories have been developed in terms of the chiral dynamics with the pseudoscalar (PS) meson degrees of freedom, since the PS mesons are the massless-mode realization of SBCS, i.e., chiral perturbation theory (ChPT) for instance~\cite{Leutwyler:1993iq}. Beyond perturbation, $p$-wave baryon resonances of negative parity have been explored with the help of unitarity condition in the coupled-channel approach~\cite{Magas:2005vu,Jido:2003cb,Jido:2002zk,Hyodo:2011ur} with much success.  

It is worth mentioning that the chiral unitary model (ChUM) is a very useful theoretical tool to address low-lying $s$-wave baryon resonances~\cite{Hyodo:2011ur}. Among successful descriptions of the resonances, the model suggests that $\Lambda(1405,1/2^-)\equiv\Lambda^*$ is a meson-baryon molecular state rather than a simple three-quark state. Moreover, the line shape of the production data via $\gamma p\to K^+\pi\Sigma$ can be interpreted by the \textit{destructive} interference of the two poles in the complex energy plane~\cite{Nam:2017yeg}, where the higher-mass pole at $(1430+15i)$ MeV couples strongly to the $\bar{K}N$ channel and the lower-mass pole at $(1376+63i)$ MeV to the $\pi\Sigma$ one by analyzing the residues of the PS-meson-baryon scattering amplitudes in ChUM.  The two poles appear in the second Riemann sheet as resonances through the attractive $S=-1$ interactions in the coupled-channel amplitudes~\cite{Hyodo:2011ur,Frazer:1964zz}. 

The pole positions were extensively investigated theoretically fitting the experimental data via ChUM~\cite{Roca:2013av,Roca:2013cca}. The two-pole scenario has been supported in various theoretical approaches, including the recent lattice-QCD simulation for its strange magnetic form factor~\cite{Hall:2014uca} and the detailed analysis for the low-energy $\bar{K}N$ amplitude~\cite{Cieply:2016jby}.

Recently, the CLAS (CEBAF Large Acceptance Spectrometer) collaboration at Jefferson laboratory (JLab) reported in Ref.~\cite{Lu:2013nza} that the invariant-mass line shape from the electro-production of kaon in $\gamma^*p\to K^+\pi\Sigma$ shows two bump structures in the vicinity of $M_{\pi\Sigma}=(1.35\sim1.45)$ GeV which is quite different from that of photo-production data~\cite{Niiyama:2009zza,Moriya:2013hwg,Moriya:2013eb}. In our previous work~\cite{Nam:2017yeg}, this distinctive features of the invariant-mass  line shape was explained by the different interference pattern due to the electromagnetic (EM) form factors of the two poles: The interference between the two poles becomes \textit{constructive}, resulting in the two bumps appearing near the higher and lower pole positions. Although there were some theoretical uncertainties, this observation supports the two-pole scenario for the $\Lambda^*$ structure~\cite{Sekihara:2008qk}.  

In the present work, we would like to investigate the interference effects of the two poles and the invariant-mass line shapes carefully, in a different production-reaction process with the vector kaon ($K^*$), i.e., $\gamma^{(*)}p\to K^{*+}\pi^0\Sigma^0$. Because there are so far only a limited experimental data for $K^*$ productions, many of our results show here are predictions and provide a guideline for the future experiments. This reaction process can be performed experimentally by the LEPS (Laser Electron Photon beam-line at SPring-8) at SPring-8 (Super Photon Ring - 8 GeV) and CLAS collaborations in the future. Because this reaction process does not contain the $\Sigma^*(1385)$ contribution, one is free from the interference between $\Sigma^*$ and $\Lambda^*$, resulting in a clear signal only from $\Lambda^*$ in the vicinity of the invariant mass $M_{\pi^0\Sigma^0}=(1.35\sim1.45)$ GeV. In addition to focusing on the differences between the photo- and electro-productions of $\Lambda^*$, one of the theoretical motivations of the present work is to estimate the strong-coupling strength of $g_{K^*N\Lambda^*}$, which is rarely studied in theories and experiments. For instance, in Ref.~\cite{Khemchandani:2011mf}, $g_{K^*N\Lambda^*}$ was estimated using the ChUM approach with the Kroll-Ruddermann (KR) interaction in terms of the vector dominance. Taking the present situation into consideration, we also want to provide a unique experimental method to determine the $K^*N\Lambda^*$ interaction strength by taking into account the incident-photon polarizations. The study of vector mesons provides information of their dynamics which is as important as of the Nambu-Goldstone bosons.

As a theoretical tool to study the present reaction process, we employ the effective Lagrangian method at the tree-level Born approximation together with the theoretical and experimental inputs for the strong and electromagnetic (EM) hadron properties. As for the strong form factors, we make use of the conventional Lorentzian type as done in our previous works~\cite{Nam:2013nfa,Nam:2017yeg}, whereas the neutral-hyperon EM form factors are parameterized by using the information of their electric charge radii, which were computed by various theoretical models including ChUM~\cite{Sekihara:2008qk,Kaskulov:2003bg}. The theoretical results of Ref.~\cite{Hawes:1998bz} are used to parameterize the EM form factor of the charged vector kaon, and the $K^*\to \gamma K$ transition form factor is devised by combining the light-cone sum rule~\cite{Khodjamirian:1997tk} and the kaon light-cone wave function, which was computed using the nonlocal chiral-quark model based on the instanton QCD vacuum by the authors~\cite{Nam:2006au,Nam:2006sx}. The gauge-invariant prescription is also taken into account to satisfy the Ward-Takahashi (WT) identity for both of the photo- and electro-productions~\cite{Nam:2017yeg}. 

As for the invariant-mass line shape $d\sigma/dM_{\pi^0\Sigma^0}$, we find that the lower-pole contribution of $\Lambda^*$ turns out to be much smaller than the higher one, when results of ChUM results are employed in reaction calculations. Therefore, we do not have considerable interferences between the two poles for $\Lambda^*$ in the present reaction process. The background (BKG) contribution comes mainly from the destructive and constructive interferences between $\Lambda$ and $\Sigma$ ground-state contributions for the photo- and electro-productions, respectively, due to that their EM  form factors change the relative signs between the invariant amplitudes. We also find that the $\Lambda^*$ peak is clearly observed for the photo- and electro-productions with the finite $K^*N\Lambda^*$ interaction, whereas the peak survives only for the photo-production when we ignore the interaction. These behaviors of the peak can be finally understood by the different $Q^2$ dependences in the $K^*$ electromagnetic and $K^*\to\gamma K$ transition form factors. Finally, we suggest a photon-polarization observable $\Sigma$, which identify the strength of the $K^*N\Lambda^*$ interaction uniquely. It turns out that the value of $\Sigma$ near the $\Lambda^*$ peak region $M_{\pi^0\Sigma^0}=1.43$ GeV becomes negative with the interaction and positive without it for $Q^2=0$, according to the different interaction structure of $K$ and $K^*$ with respect to the polarized photon. As for $Q^2\ne0$, we observe similar but clearer signals, but the $\Sigma$ curves slightly change, because the scalar component of the photon polarization enhances the spin-$1$ exchange.  

The present paper is organized as follows: In Section II, we briefly introduce the theoretical framework. The numerical results and relevant discussions are given in Section III, and Section IV is devoted to the summary of the present work.
\section{Theoretical framework}
\begin{figure}[t]
\includegraphics[width=12cm]{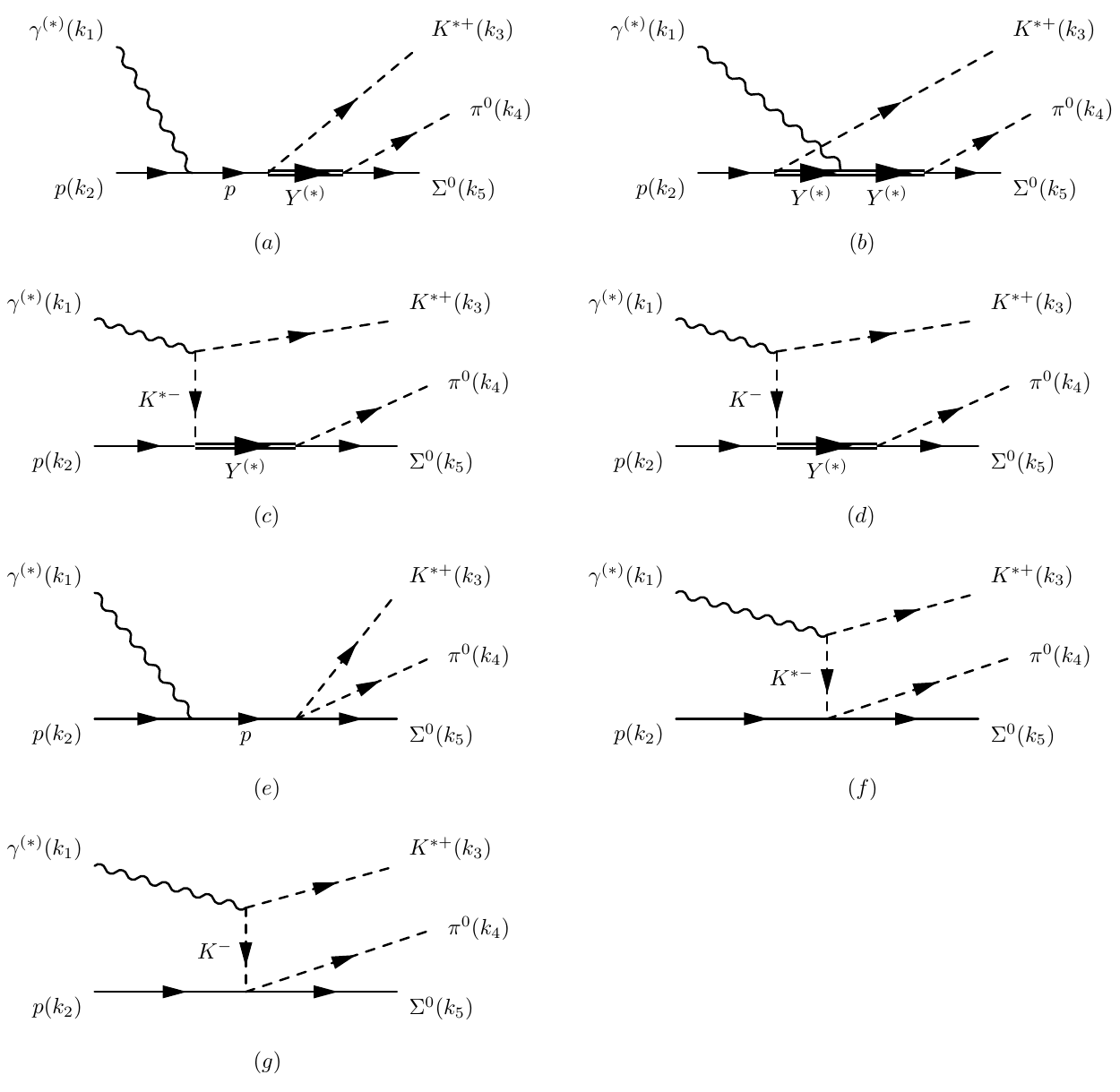}
\caption{Relevant Feynman diagrams for $\gamma p\to K^{*+}\pi^0\Sigma^0$: ($a$) proton-pole diagram in the $s$ channel, ($b$) hyperon ground-state and resonance diagram in the $u$ channel, ($c$) $K^*$-exchange in the $t$ channel, ($d$) $K$-exchange in the $t$ channel, ($e$ and $f$) background contributions from the $PB\to VB$ Kroll-Ruddermann (KR) interaction, and ($g$) those from the $PB\to PB$ Weinberg-Tomozawa (WT) interaction. As for the hyperons $Y^{(*)}$, we consider $\Lambda(1116,1/2^+)$, $\Sigma^0(1192,1/2^+)$, $H(1430,1/2^-)$, and $L(1390,1/2^-)$, in which $H$ and $L$ indicate the higher- and lower-pole contributions for $\Lambda(1405,1/2^-)$, respectively. See the text for details.}       
\label{FIG0}
\end{figure}

In this Section, we would like to describe the theoretical framework briefly, based on the effective
Lagrangian approach in the tree-level Born approximation. The relevant Feynman diagrams for $\gamma^{(*)}p\to K^{*+}\pi\Sigma$ are shown in Fig.~\ref{FIG0}, in which we also define the four momenta for each particle involved. Namely, the momenta $k_ {1\sim5}$ are for the particles participating the reaction as shown in Fig.~\ref{FIG0}. The diagrams $(a)$ and $(c)$ are responsible to satisfy the Ward-Takahashi identity (WTI) for the pseudoscalar (PS) $PBB$ coupling scheme, where $P$ and $B$ indicate the PS meson and spin-$1/2$ baryon, respectively. The diagram ($d$) denotes the PS kaon exchange. Note that the diagrams $(b)$ and $(d)$ are gauge-invariant by themselves, due to their magnetic photon-coupling structures.  As for the diagrams, where hyperons $Y^{(*)}$ appear as intermediate states, we consider $\Lambda(1116,1/2^+)$, $\Sigma^0(1192,1/2^+)$, $H(1430,1/2^-)$, and $L(1390,1/2^-)$, in which $H$ and $L$ indicate the higher- and lower-pole contributions for $\Lambda(1405,1/2^-)$, respectively. 

Note that $\Sigma^{*0}(1385)$ is not taken into account here, since it does not couples to the neutral  $\pi^0\Sigma^0$ channel, corresponding to the zero Clebsch-Gordan coefficient. However, we can consider the EM transition of $\Lambda$-$\Sigma^{*0}(1385)$, which can be shown in the diagram $(b)$ with the $\pi^0\Sigma^0\Lambda$ interaction vertex. Although the sizable transition magnetic moment is given by $\mu_{\Sigma^{*0}\to\gamma\Lambda}=(2.75\pm0.25)\,\mu_N$ from the experimental analysis~\cite{Keller:2011nt} and $2.28\,\mu_N$ from the quark model~\cite{Dhir:2009ax}, since this contribution plays the role of backgrounds and is suppressed more than the other contributions due to the $u$-channel strong form factor near the threshold~\cite{Nam:2017yeg}. Moreover, we do not have much information on the transition form factor as a function of $Q^2$. Therefore, we do not consider this contribution explicitly, reducing theoretical uncertainties. 

In addition, there can be the contributions from nucleon resonances with $M_{N^*}\approx2$ GeV in the diagram $(a)$ by interchanging $\pi^0$ and $K^{*+}$ for instance~\cite{Wang:2017tpe,Huang:2018qym}. In general, the $N^*$ contributions interfere with the hyperon resonances, such as $\Lambda(1405)$. As observed in the $\gamma p \to K^+K^-p$ reaction process~\cite{Ryu:2016jmv}, such interference effects are not significant and becomes negligible, when the decay width of the resonance is considerably wide. Moreover, the coupling constants for $K^*N^*\Sigma$ are rarely determined experimentally, resulting in the increase of theoretical uncertainties. Hence, we do not consider the $N^*$ contributions in the present work rather safely. 

The diagrams are derived from the Kroll-Ruderman (KR) $VB\to PB$ ($e$ and $f$) and Weinberg-Tomozawa (WT) $PB\to PB$ ($g$) interactions~\cite{Khemchandani:2011mf}. Here, $V$ stands for the vector mesons, such as $K^*(892,1^-)$. We note that there are more diagrams, in which the photon couples to the outgoing $\Sigma$. We, however, verified that their contributions are much smaller than those from the diagrams shown in Fig.~\ref{FIG0}. Therefore, we ignore them in the present work.   

To compute the diagrams in Fig.~\ref{FIG0}, we define the  effective Lagrangians for the electromagnetic (EM) and strong interaction vertices as follows:
\begin{eqnarray}
\label{eq:EFFLAG}
\mathcal{L}_{\gamma BB}&=&
-\bar{B} \left[ e_B\rlap{\,/}{A} - \frac{e\kappa_B }{4M_N} 
(\sigma\cdot F) \right] B\,\,\,\mathrm{for}\,\,\,B=N,\Lambda,\Sigma,H,L,
\cr
\mathcal{L}_{\gamma K^*K} &=&
g_{\gamma KK^*} \varepsilon^{\mu\nu\alpha\beta} \left( \partial_\mu A_\nu \right) 
\left(\partial_\alpha K_\beta^* \right) K+\mathrm{h.c.},
\cr
\mathcal{L}_{\gamma K^*K^*} &=&
-ie_{K^*}A^\mu (K^{*\nu}\mathcal{K}_{\mu\nu}^{*\dagger}
-\mathcal{K}_{\mu\nu}^{*}K^{*\dagger\nu}),\,\,\,\,\mathcal{K}_{\mu\nu}=\partial_\mu K^*_\nu-\partial_\nu K^*_\mu,
\cr
\mathcal{L}_{K^* N(H,L)} &=&
-g^V_{K^* N(H,L)} (\bar{H},\bar{L})  \rlap{\,/}{K}^*\gamma_5 N + \mathrm{h.c.}, 
\cr
\mathcal{L}_{\pi\Sigma (H,L)} &=&-ig_{\pi\Sigma (H,L)}(\bar{H},\bar{L})
\bm{\pi}\cdot\bm{\Sigma}+\mathrm{h.c.}, 
\cr
\mathcal{L}_{KN(H,L)} &=&-ig_{KN(H,L)}(\bar{H},\bar{L})
KN+\mathrm{h.c.}, 
\cr
\mathcal{L}_\mathrm{WT} &=&-ig_\mathrm{WT}\bar{\Sigma}(\pi^\dagger\rlap{/}{\partial}K
-K^\dagger\rlap{/}{\partial}\pi)N+\mathrm{h.c.},
\cr
\mathcal{L}_\mathrm{KR} &=&-iG_\mathrm{KR}g_\mathrm{KR}
\bar{B}\gamma^\mu\gamma_5(P^\dagger K^*_\mu-PK^{*\dagger }_\mu) B+\mathrm{h.c.}.
\end{eqnarray}
As for the $\pi\Sigma B$ and $KNB$ interactions we replace $(H,L)$ into $\gamma_5B$ in the Lagrangians. Due to the lack of the experimental and theoretical information, we ignored the tensor coupling $g^T_{K^*N(H,L)}$ throughout this work. The $PBB$ coupling constants corresponding to $H$ and $L$ were estimated by the residues for the higher and lower poles for the $\Lambda(1405)$ in the complex energy plane~\cite{Hyodo:2011ur}. In Ref.~\cite{Khemchandani:2011mf}, the values for $g_{K^*N(H,L)}$ and $g_{KN(H,L)}$ were given as functions of the KR coupling $g_\mathrm{KR}\equiv g=(1\sim6)$. Since the values of $g_{K^*N(H,L)}$ are nearly proportional to $g$~\cite{Khemchandani:2011mf}, we examine the case with $g=1$ for the most part of the present work. As for the charged transition $\gamma PV$ vertex, we use $g_{\gamma K^{*+}K^+}=0.254\,\mathrm{GeV}^{-1}$~\cite{Olive:2016xmw}.

In addition, we will take the $\Lambda(1116,1/2^+)$ and $\Sigma(1192,1/2^+)$ contributions for the diagram $(b)$ as the hyperon backgrounds (BKG). The strong-coupling constants for $\Lambda$ and $\Sigma$ are employed from the Nijmegen soft-core potential model (NSC97)~\cite{Stoks:1999bz}. We also take the WT and KR contact interactions into account as the BKG contributions. The coupling constants for those contributions are given by $g_\mathrm{WT}=1/8f^2_\pi$ and $G_\mathrm{KR}=(D-F)/2f_\pi$ for the $\pi^0\Sigma^0$ channels with $D-F=0.34$ and $f_\pi=93.3$ MeV~\cite{Jido:2002zk}. All the values for the relevant couplings are listed in Table.~\ref{TAB1}. 

\begin{table}[b]
\begin{tabular}{c|cccccc}
&$g_{KN \Lambda_{H}}$&$g_{K^* N\Lambda_{H}}$&$g_{\pi\Sigma \Lambda_{H}}$
&$g_{KN \Lambda_{L}}$&$g_{K^* N\Lambda_{L}}$&$g_{\pi\Sigma \Lambda_{L}}$\\
\hline
$g=0$&$2.4+1.1i$&$0.0+i0.0$&$-0.2-1.4i$&$1.4-1.6i$&$0.0+0.0i$&$-2.3+1.4i$\\
\hline$g=1$
&$2.4+1.1i$&$0.1-0.9i$&$-0.2-1.3i$&$1.4-1.6i$&$-0.4+0.1i$&$-2.3+1.5i$\\
\hline
\end{tabular}
\begin{tabular}{ccccccccccc}
$g_{KN \Lambda}$&$g_{K^* N\Lambda}$&$g_{\pi\Sigma \Lambda}$&$g_{KN \Sigma}$&
$g_{K^* N\Sigma}$&$g_{\pi\Sigma\Sigma}$&$g_{K^* N\Sigma^*}$&$g_{\pi\Sigma\Sigma^*}$
&$g_{\gamma K^{*+}K^-}$&$g_\mathrm{WT}$&$G_\mathrm{KR}$\\
\hline
$-13.89$&$-4.26$&$11.86$&$4.08$&$-2.46$&$11.89$&$-2.60$
&$0.55$&$-\frac{0.254}{\mathrm{GeV}}$
&$\frac{1}{8f^2_\pi}$&$\frac{0.34}{4f_\pi}$\\
\end{tabular}
\caption{Relevant coupling constants for the present work. Note $g=g_{KR}$ defined in Ref~\cite{Khemchandani:2011mf}.}
\label{TAB1}
\end{table}

The invariant amplitudes, shown in the diagrams $(a\sim d)$, for the \textit{resonance} $(H,L)$ contributions are computed from the effective Lagrangians and resulted in
\begin{eqnarray}
\label{eq:AMP1}
i\mathcal{M}^{H,L}_{a}&=&
\frac{g_{H,L}F_cF_\mathrm{EM}^p\bar{u}_\Sigma(\rlap{/}{q}_{4+5}+M_{H,L})
\rlap{/}{\varepsilon}^*\gamma_5(\rlap{/}{q}_{1+2}+M_N)
\rlap{/}{\Gamma}_a(Q^2)u_N}
{[q^2_{4+5}-M^2_{H,L}-iM_{H,L}\Gamma_{H,L}]
[q^2_{1+2}-M^2_N]},
\cr
i\mathcal{M}^{H,L}_{b}&=&
\frac{g_{H,L}F_uF_\mathrm{EM}^{H,L}\bar{u}_\Sigma(\rlap{/}{q}_{4+5}+M_{H,L})
\rlap{/}{\Gamma}_b(Q^2)
(\rlap{/}{q}_{2-3}+M_{H,L})\rlap{/}{\varepsilon}^*\gamma_5u_N}
{[q^2_{4+5}-M^2_{H,L}-iM_{H,L}\Gamma_{H,L}]
[q^2_{2-3}-M^2_{H,L}]},
\cr
i\mathcal{M}^{H,L}_{c}&=&
\frac{g_{H,L}F_cF_\mathrm{EM}^{K^*}\bar{u}_\Sigma(\rlap{/}{q}_{4+5}+M_{H,L})
\rlap{/}{\varepsilon}^*\gamma_5\Gamma_c(Q^2)u_N}
{[q^2_{4+5}-M^2_{H,L}-iM_{H,L}\Gamma_{H,L}]
[q^2_{1-3}-M^2_{K^*}]},
\cr
i\mathcal{M}^{H,L}_{d}&=&-
\frac{ig'_{H,L}F^K_tF_\mathrm{EM}^{K\to K^*}\bar{u}_\Sigma(\rlap{/}{q}_{4+5}+M_{H,L})
(\epsilon^{\mu\nu\alpha\beta}k_{1\mu}\epsilon_\nu k_{3\alpha}\varepsilon^*_\beta)u_N}
{[q^2_{4+5}-M^2_{H,L}-iM_{H,L}\Gamma_{H,L}]
[q^2_{1-3}-M^2_K]},
\end{eqnarray}
where $q_{i\pm j}=k_i\pm k_j$. The polarization vectors for the incident photon and outgoing $K^*$ are denoted by $\epsilon_\mu$ and $\varepsilon_\mu$, respectively. We used the combined coupling constants $g_{H,L}=e_{K^*}g_{K^*N(H,L)}g_{\pi\Sigma(H,L)}$ and $g'_{H,L}=g_{\gamma KK^*}g_{KN(H,L)}g_{\pi\Sigma(H,L)}$ for convenience. The vertex functions $\Gamma_{a,b,c}$ are given by
\begin{eqnarray}
\label{eq:GAMMASTR}
\rlap{/}{\Gamma}_a(Q^2)&=&\rlap{/}{\epsilon}+(F^p_1-1)\left[\rlap{/}{\epsilon}+\frac{(\epsilon\cdot k_1)\rlap{/}{k}_1}{Q^2}\right]
-\frac{\kappa_N F^p_2}{4M_N}(\rlap{/}{k}_1\rlap{/}{\epsilon}
-\rlap{/}{\epsilon}\rlap{/}{k}_1),
\cr
\rlap{/}{\Gamma}_b(Q^2)&=&
F^{H,L}_1\left[\rlap{/}{\epsilon}+\frac{(\epsilon\cdot k_1)\rlap{/}{k}_1}{Q^2}\right]
-\frac{\kappa_{H,L} F^{H,L}_2}{4M_N}(\rlap{/}{k}_1\rlap{/}{\epsilon}
-\rlap{/}{\epsilon}\rlap{/}{k}_1),
\cr
\Gamma_c(Q^2)&=&\epsilon\cdot (2k_3-k_1)+(F^{K^*}-1)
\left[\epsilon\cdot (2k_3-k_1)+\frac{(\epsilon\cdot k_1)[\epsilon\cdot (2k_3-k_1)]}{Q^2}\right].
\end{eqnarray}

The phenomenological prescription for the vertices in Eq.~(\ref{eq:GAMMASTR}) satisfies the WTI for $Q^2=0$~\cite{Davidson:2001rk,Nam:2013nfa} and $Q^2\ne0$~\cite{Nam:2017yeg} simultaneously, and the invariant amplitudes defined in Eq.~(\ref{eq:AMP1}) can be also used similarly for the $\Lambda$ and $\Sigma$ BKG contributions by changing the baryon field $(H,L)$ into $\gamma_5B$. Note that the invariant amplitude $\mathcal{M}_{d}$ is gauge-invariant by itself, due to the anti-symmetric tensor-coupling structure. Therefore, we multiply $F^K_t$, not $F_c$. 

In Eq.~(\ref{eq:AMP1}), we have introduced the strong ($F_{x=s,t,u}$) and electromagnetic ($F_\mathrm{EM}$) form factors to reproduce experimental data, considering the internal structure of the hadrons. As for the strong form factors, the following parameterized form is employed:
\begin{equation}
\label{eq:SFF}
F_x=\frac{\Lambda^4_\mathrm{strong}}{\Lambda^4_\mathrm{strong}+(x-M_x)^2},
\,\,\,\,
F_c=1-(1-F_s)(1-F^{K^*}_t).
\end{equation}
Here, $\Lambda_\mathrm{strong}$ indicates the strong cutoff and the Mandelstam variables $x$ are defined as $s=(k_1+k_2)^2$, $t=(k_1-k_3)^2$, and $u=(k_2-k_3)^2$. Note that the invariant amplitudes $i\mathcal{M}_{b}$ and $i\mathcal{M}_{d}$ do not contain the common strong form factor $F_c$, since they are gauge-invariant by themselves. $M_x$ stands for the mass of the propagating particles: $M_{s,t,u}=M_{K,K^*,Y}$. As for the $K$-exchange diagram ($d$) in Fig.~\ref{FIG0}, we assign the strong form factor $F^K_t$ with $M_x=M_K$. 

Here are some discussions for the prescription for the strong form factors. In principle, one can consider different form factors for each amplitude as functions of Mandelstam variables. However, if this is the case, additional amplitudes, which can not be read from Feynman diagrams, are necessary to satisfy WTI. In order to avoid this theoretical uncertainty, the authors of Ref.~\cite{Davidson:2001rk}  suggested to multiply a common form factor ($F_c$) to the sum of the gauge-invariant combinations of bare amplitudes, and the individual one ($F^K_t$ and $F_u$) to the the amplitudes which are gauge invariant by themselves. 

The Dirac ($F_1$) and Pauli ($F_2$) EM form factors for the proton read with its anomalous magnetic moment $\kappa_p=1.79$:
\begin{equation}
\label{eq:DFFF}
F^p_1(Q^2)=\frac{G^p_E(Q^2)+\tau G^p_M(Q^2)}{1+\tau},
\,\,F^p_2(Q^2)=\frac{G^p_M(Q^2)-G^p_E(Q^2)}{\kappa_B(1+\tau)},\,\,\tau=\frac{Q^2}{4M^2_p},
\end{equation}
where the Sachs form factors are defined by~\cite{Perdrisat:2006hj}
\begin{equation}
\label{eq:SACHS}
G^p_E(Q^2)=G_D(Q^2),\,\,\,\,G^p_M(Q^2)=(\kappa_B+1)G_D(Q^2),\,\,\,\,
G_D(Q^2)=\left(\frac{1}{1+Q^2/\Lambda^2_D}\right)^2,\,\,\,\,\Lambda^2_D=0.71\,\mathrm{GeV^2}.
\end{equation}
As for the neutral hyperons ($Y$) such as $\Lambda(1116)$, $\Sigma(1193)$, $H$, and $L$, we
make use of the following form-factor parameterization, which was employed to reproduce the data for $\gamma^*p\to K^+\pi\Sigma$~\cite{Kaskulov:2003bg}:
\begin{eqnarray}
\label{eq:NEUTEMFF}
G^Y_E(Q^2)=-\frac{\langle r^2_E\rangle_Y}{6}Q^2F_K(Q^2)\left(\frac{1}{1+Q^2\langle r^2_M\rangle_Y/12}\right)^2G_D(Q^2),\,\,\,\,
G^Y_M(Q^2)=\mu\left(\frac{1}{1+Q^2\langle r^2_M\rangle_Y/12}\right)^2G_D(Q^2).
\end{eqnarray}
To obtain the Dirac and Pauli form factors for those hyperons, the following expressions are used: 
\begin{equation}
\label{eq:DFFFs}
F^Y_1(Q^2)=\frac{[G^Y_E(Q^2)+\tau G^Y_M(Q^2)]}{1+\tau},
\,\,F^Y_2(Q^2)=\frac{[G^Y_M(Q^2)-G^Y_E(Q^2)]}{\kappa_Y(1+\tau)}.
\end{equation}
We emphasize that the Sachs form factors in this parameterization of Eq.~(\ref{eq:NEUTEMFF}) are defined with the electric and magnetic mean-square charge radii $\langle r^2_{E,M}\rangle_Y$. Although there are various theoretical estimations for $\Lambda(1116)$ and $\Sigma(1193)$, we use $\langle r^2_{E}\rangle=0.029\,\mathrm{fm}^2$  and $0.209\,\mathrm{fm}^2$, respectively, from the relativistic quark model~\cite{Berger:2004yi}, and the values of $\langle r^2_{M}\rangle$ are chosen to be zero for simplicity. We verified that the finite values of $\langle r^2_{M}\rangle$ do not make qualitative differences in the corresponding form factors as far as we resort to the parameterization in Eq.~(\ref{eq:NEUTEMFF}).

From the ChUM calculations~\cite{Sekihara:2008qk}, the values of $\langle r^2_{E,M}\rangle_{H,L}$ were estimated as listed in Table~\ref{TAB2} and they will be used to obtain the EM form factors  for $H$ and $L$. We note that the complex values of $\langle r^2_{E,M}\rangle_{H,L}$ will provide additional phase factors to the amplitude of  electro-production. The anomalous magnetic moments for the hyperons are chosen to be $\kappa_{\Lambda,\Sigma,H,L}=(-0.64,0.72,0.40,0.30)$ from experiments and theories~\cite{Olive:2016xmw,Sekihara:2008qk}. 
\begin{table}[b]
\begin{tabular}{cccc}
$\langle r^2_{E}\rangle_H$&$\langle r^2_M\rangle_H$&$\langle r^2_{E}\rangle_L$&$\langle r^2_{M}\rangle_L$\\
\hline
$-0.131+0.303i$&$0.267-0.407i$&$0.018+0.002i$&$-0.013+0.021i$
\end{tabular}
\caption{The electric and magnetic mean-square charge radii for the high ($H$) and low ($L$) pole contributions [$\mathrm{fm}^2$].}
\label{TAB2}
\end{table}

\begin{figure}[t]
\begin{tabular}{cc}
\includegraphics[width=5.5cm]{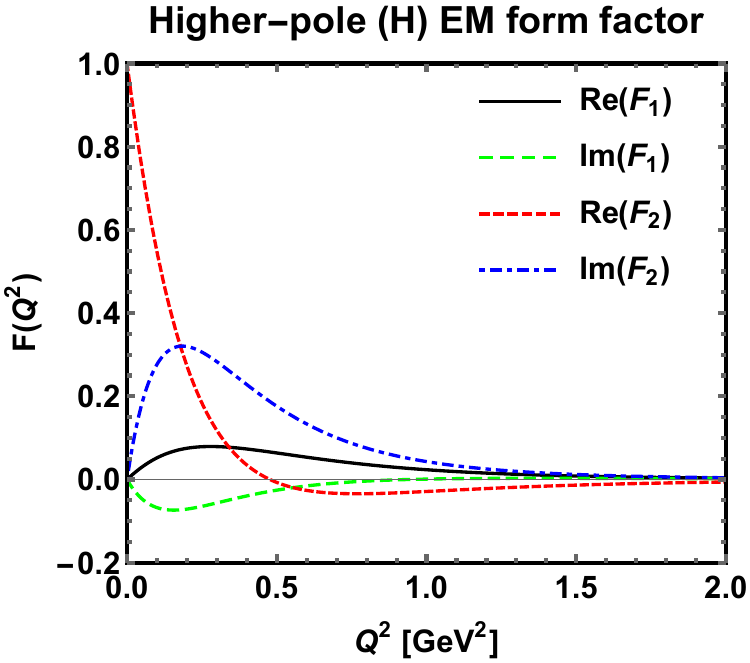}
\includegraphics[width=5.5cm]{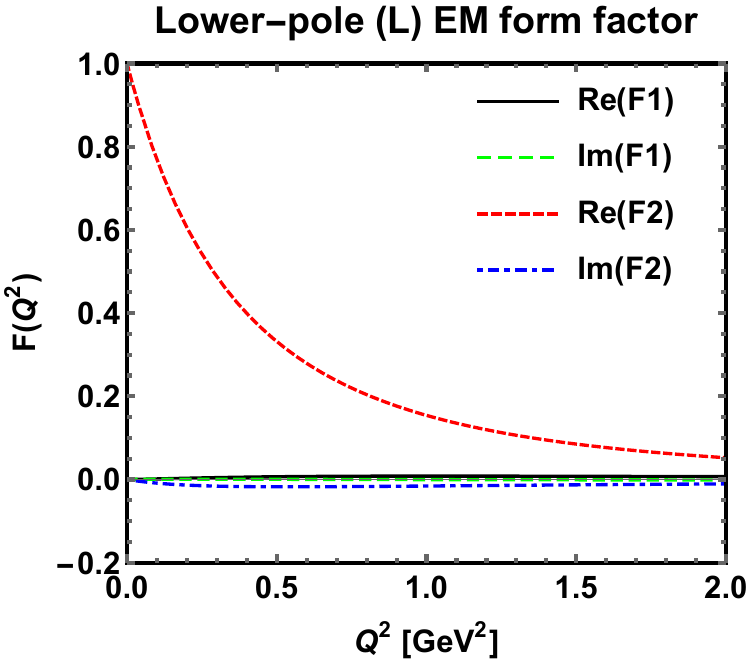}
\includegraphics[width=5.5cm]{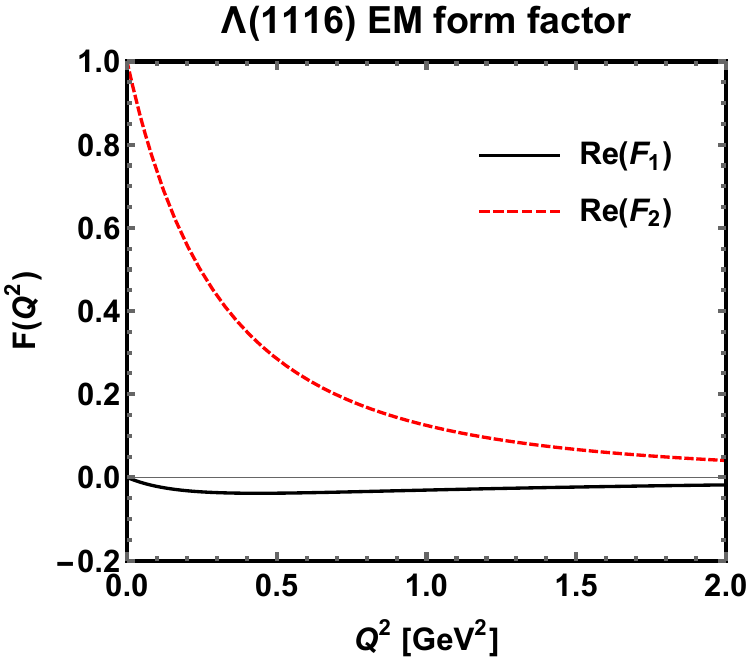}
\end{tabular}
\begin{tabular}{cc}
\includegraphics[width=5.5cm]{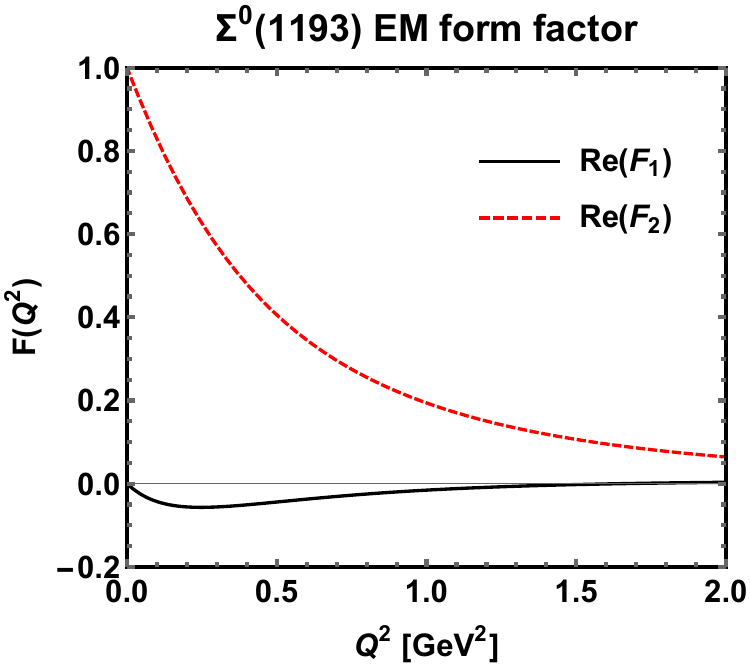}
\includegraphics[width=5.5cm]{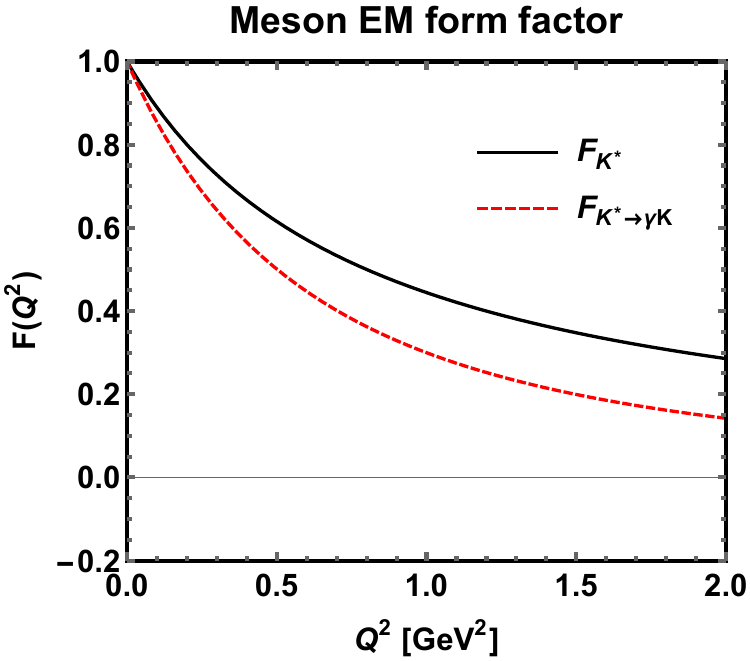}
\end{tabular}
\caption{(Color online) EM form factors for the neutral hyperons as functions of $Q^2$ [GeV$^2$]. The solid, dashed, dotted, and dot-dashed lines indicate $\mathrm{Re}F_1$, $\mathrm{Im}F_1$, $\mathrm{Re}F_2$, and $\mathrm{Im}F_2$, respectively. Here, $F_1$ and $F_2$ stand for the Dirac and Pauli form factors for spin-$1/2$ baryons.}       
\label{FIG1}
\end{figure}

The charged vector-kaon EM form factor is parameterized with its mean-square charge radius by 
\begin{equation}
\label{eq:FKEM}
F^{K^*}_\mathrm{EM}(Q^2)=\frac{1}{1+Q^2\langle r^2\rangle_{K^*}/6},
\end{equation}
and $\langle r^2\rangle_{K^*}=0.54\,\mathrm{fm}^2$, calculated from the Lorentz-covariant Dyson-Schwinger method~\cite{Hawes:1998bz}. As for the EM transition vertex with $K^*\to \gamma K$,  $F^{K^*\to\gamma K}_\mathrm{EM}(Q^2)$ is necessary and defined by a simple flavor-SU(3)-symmetric extension of the result given in Ref.~\cite{Khodjamirian:1997tk}:
\begin{equation}
\label{eq:TRFF1}
F^{K^*\to\gamma K}_\mathrm{EM}(Q^2)=\frac{F_K}{3F_{K^*}}V(Q^2,M^2),\,\,\,\,
V(Q^2,M^2)\approx\int^1_0\frac{d\mathrm{u}}{\mathrm{u}}
\phi_K(\mathrm{u})\,
\exp\left[-\frac{Q^2(1-\mathrm{u})}{\mathrm{u}\Lambda^2}+\frac{m^2_{K^*}}{\Lambda^2}\right],
\end{equation}
where $\phi_K(\mathrm{u})$ denotes the kaon light-cone wave function as a function of the longitudinal momentum fraction $\mathrm{u}=(0\sim1)$ for a quark inside the meson. The value of $\Lambda$ in Eq.~(\ref{eq:TRFF1}) indicates the nonperturbative scale at which $\phi_K(\mathrm{u})$ is evaluated. In Refs.~\cite{Nam:2006au,Nam:2006sx}, the present authors employed the nonlocal chiral-quark model to compute $\phi_K(\mathrm{u})$, being based on the instanton model, resulting in the following Gegenbauer polynomial expression:
\begin{equation}
\label{eq:LCWF}
\phi_K(\mathrm{u})\approx6u(1-\mathrm{u})[1+3a^K_1(2\mathrm{u}-1)],
\end{equation}
where the asymmetry coefficient turns out to be $a_1^K=0.06865$ at the energy scale of $\Lambda=1.2$ GeV with the empirical values $f_{K,K^*}=(156.1,204)$ MeV. Since we are also interested in the present reaction process in the low-energy region near the threshold, it must be consistent for us to use the above $\phi_K(u)$ for the transition form factor here. Using Eqs.~(\ref{eq:TRFF1}) and (\ref{eq:LCWF}), we introduce the following dipole-type parameterization for numerical convenience
\begin{equation}
\label{eq:TRFF2}
F^{K^*\to\gamma K}_\mathrm{EM}(Q^2)=\left(\frac{\Lambda^2_{K^*\to\gamma K}}{\Lambda^2_{K^*\to\gamma K}+Q^2}\right)^2
,\,\,\,\,\Lambda_{K^*\to\gamma K}=1.1\,\mathrm{GeV}.
\end{equation}
All the $Q^2$ dependence of the relevant EM form factors are plotted in Fig.~\ref{FIG1}. Note that we observe a tendency that $F^{K^*}_\mathrm{EM}\ge F^{K^*\to \gamma K}_\mathrm{EM}$ for all the $Q^2$ regions. In the numerical results, this tendency will play an important role in making the invariant-mass line shapes for the electro-production with and without the $K^*N\Lambda^*$ interaction. 

Although we have used the PS coupling scheme for the $PBB$ and $PB(H,L)$ Yukawa vertices, as shown in Eq.~(\ref{eq:EFFLAG}), there can be the pseudo-vector (PV) coupling:
\begin{equation}
\label{eq:PSPV}
\mathcal{L}^\mathrm{PV}_{PBB'}=-\frac{g^\mathrm{PV}_{PBB'}}{2f_\pi}\bar{B}'(\Gamma_5\rlap{/}{\partial}P)B+\mathrm{h.c.},
\end{equation}
where $\Gamma_5=(\gamma_5,1_{4\times4})$ for the positive- or negative-parity $B$, whereas $B'$ is assumed to have positive parity. The PS and PV couplings satisfy the Goldberger-Treiman (GT) relations~\cite{Nam:2003uf} by
\begin{equation}
\label{eq:GT}
g^\mathrm{PV}_{PBB'}
=\frac{2f_\pi g^\mathrm{PS}_{PBB'}}{|\Pi M+\Pi'M'|},
\end{equation}
where $\Pi$ and $M$ indicate the parity and mass of the corresponding baryon, respectively. In the PV scheme, we may have an additional term corresponding to the $\gamma PBB'$ contact interaction to preserve the WTI. However, in the present reaction process, the contact interaction disappears, because of the neutral pion electric charge. Using the PV Lagrangian and GT relation in Eqs.~(\ref{eq:PSPV}) and (\ref{eq:GT}), the corresponding invariant amplitudes and coupling constants can be computed in a straightforward manner. However, we have verified that the the PV scheme gives essentially the same results as the PS scheme by tuning the cutoff parameter for the form factors, defined in Eq.~(\ref{eq:SFF}), by about $(10\sim15)\%$. Hence, according to this observation, we will show the numerical results only from the PS scheme hereafter.

There are additional BKG contributions from the KR and WT interactions as shown in the diagrams ($e\sim g$) and their invariant amplitudes satisfying the WTI read:
\begin{eqnarray}
\label{eq:WTKR}
i\mathcal{M}^\mathrm{KR}_{e}&=&
\frac{eG_\mathrm{KR}F^{K^*}_tF^{K^*}_\mathrm{EM}\bar{u}_\Sigma\gamma_5
\rlap{/}{\varepsilon}^*(\rlap{/}{q}_{1+2}+M_N)
\rlap{/}{\Gamma}_a(Q^2)u_N}{[q^2_{1+2}-M^2_N]},
\cr
i\mathcal{M}^\mathrm{KR}_{f}&=&-
\frac{eG_\mathrm{KR}F^{K^*}_tF^{K^*}_\mathrm{EM}\bar{u}_\Sigma\gamma_5
\rlap{/}{\varepsilon}^*\Gamma_b(Q^2)u_N}
{[q^2_{1-3}-M^2_{K^*}]},
\cr
i\mathcal{M}^\mathrm{WT}_{g}&=&
-\frac{ig_\mathrm{WT}g_{\gamma KK^*}
F^K_tF^{K^*\to \gamma K}\bar{u}_\Sigma(\rlap{/}{k}_1-\rlap{/}{k}_3+\rlap{/}{k}_4)
(\epsilon^{\mu\nu\alpha\beta}k_{1\mu}\epsilon_\nu k_{3\alpha}\varepsilon^*_\beta)u_N}
{[q^2_{1-3}-M^2_K]}.
\end{eqnarray}
We verify that these KR and WT contributions are numerically very tiny in comparison to the others, due to the small values of $g_\mathrm{WT}$ and $G_\mathrm{KR}$ for $g=1\sim6$.

The total amplitude for the present reaction process can be written with the resonance and BKG contributions as follows:
\begin{equation}
\label{eq:TAMP}
i\mathcal{M}_\mathrm{total}=\underbrace{e^{i\phi}(i\mathcal{M}_H+i\mathcal{M}_L)}_\mathrm{Resonance}
+\underbrace{i\mathcal{M}_\Lambda+i\mathcal{M}_\Sigma+i\mathcal{M}_\mathrm{WT}+i\mathcal{M}_\mathrm{KR}.}_\mathrm{BKG}
\end{equation}
Note that the phase factor between the $\Lambda$ and $\Sigma$ contributions are determined by the Nijmegen potential model. Although there can be certain phase factors between the hyperon contributions and the (KR, WT) ones, we ignore them, by taking into account the numerical results showing $|i\mathcal{M}_{\Lambda,\Sigma}|\gg |i\mathcal{M}_\mathrm{WT,KR}|$. However, since the strengths of the resonance contributions are compatible with those of the hyperon BKG ones, we introduce a phase factor $e^{i\phi}$ between them, whereas the phase between the $H$ and $L$ contributions are determined by ChUM. The phase angle $\phi$ will be treated as a free parameter $(0\le\phi\le\pi)$.

\section{Numerical results and Discussions}
\begin{table}[b]
\begin{tabular}{cccc}
$M_H$&$\Gamma_H$&$M_L$&$\Gamma_L$\\
\hline
$1430$ MeV&$30$ MeV&$1376$ MeV&$126$ MeV\\
\end{tabular}
\label{TAB3}
\caption{Input values for the masses and full-decay widths for the higher- $(H)$ and lower-pole $(L)$ contributions from the ChUM calculation~\cite{Jido:2003cb}.}
\end{table}

In this Section, we demonstrate the numerical results with relevant discussions in detail. The input values for their masses and full decay widths for $(H,L)$ are listed in Table~\ref{TAB3}, being based on the ChUM results~\cite{Khemchandani:2011mf}. We note that the masses and full-decay widths depend on the regularization schemes in general in ChUM~\cite{Nam:2003ch}, providing about a few percent differences. However, those differences do not make any considerable changes in our conclusion of the present work. 

\begin{figure}[t]
\begin{tabular}{cc}
\includegraphics[width=8.5cm]{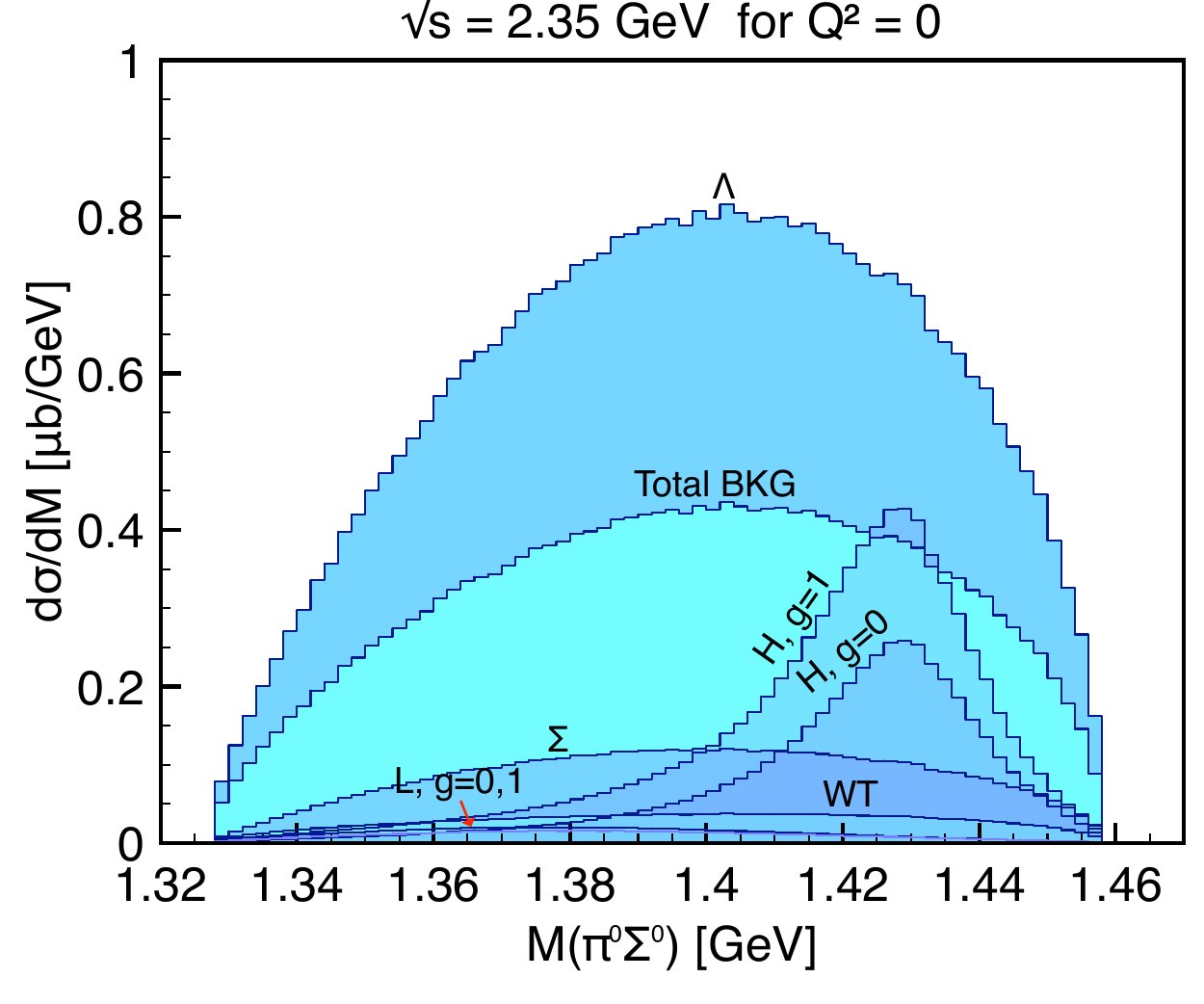}
\includegraphics[width=8.5cm]{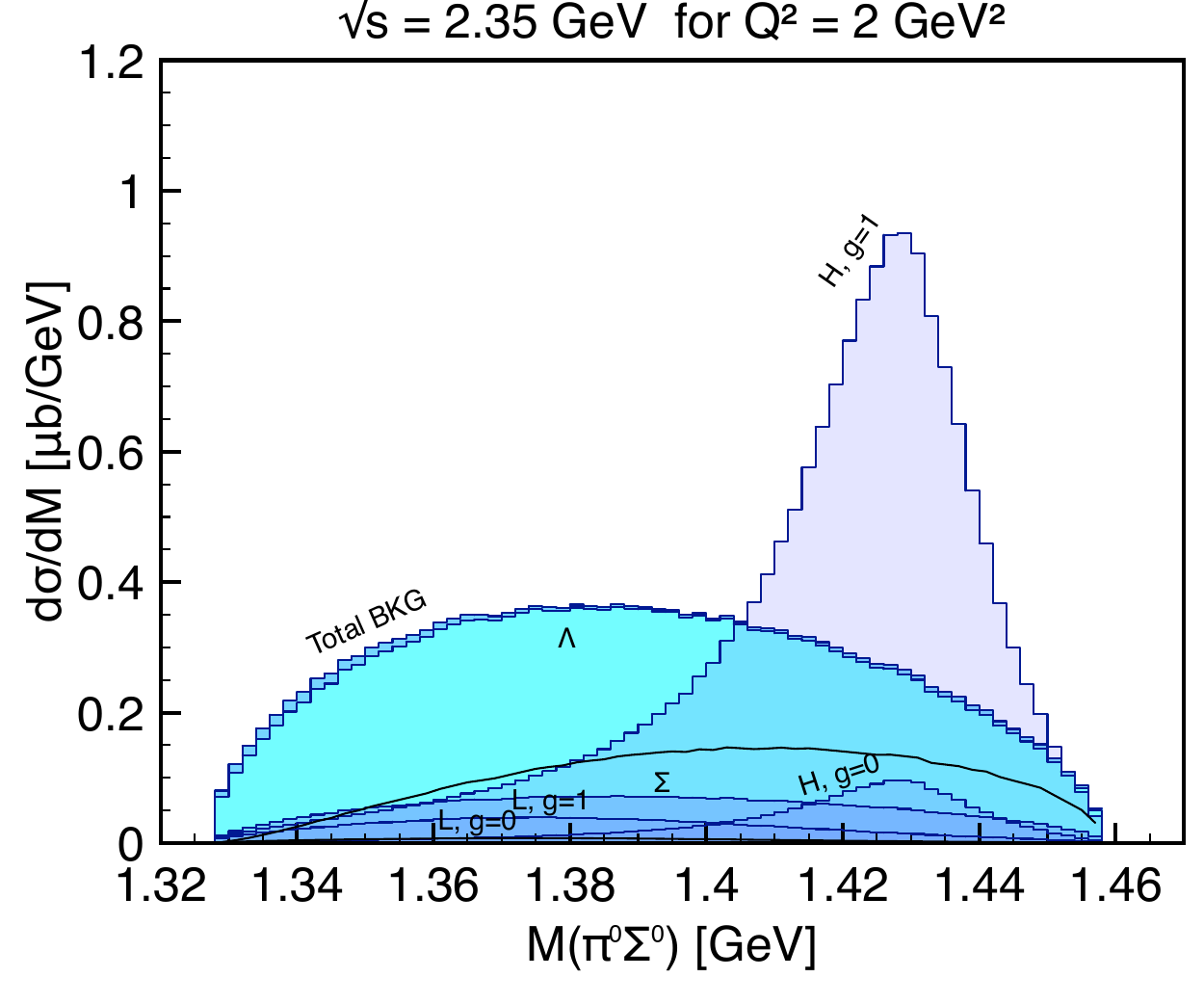}
\end{tabular}
\caption{(Color online) Each contribution for $\pi^0\Sigma^0$-invariant-mass line shape $(d\sigma/dM_{\pi^0\Sigma^0})$ for $Q^2=0$ (left) and $Q^2=2\,\mathrm{GeV}^2$ (ight) at $\sqrt{s}=2.35$ GeV.}       
\label{FIG2}
\end{figure}

First, we show the numerical results for the line shapes of invariant $M_{\pi^0\Sigma^0}\equiv M_I$ mass, i.e., $d\sigma/dM_I$, for each contribution ($H$, $L$, $\Lambda$, $\Sigma$, WT, and KR) for $Q^2=0$ (photo-production) and $Q^2=2\,\mathrm{GeV}^2$ (electro-production) in the left and right panels of Fig.~\ref{FIG2}, respectively. Here, we choose $\sqrt{s}=2.35$ GeV and the strong cutoff parameter for Eq.~(\ref{eq:SFF}) is set to be $0.9$ GeV, which was employed to reproduce the experimental data for $\gamma p\to K^+\Lambda(1405)$~\cite{Nam:2017yeg}, since we have not had experimental data for the $K^*$ production to compare with the theory at this moment. Note that the order of the cross sections for the electro-production is much smaller than those for the photo-production, due to the EM form factors as shown in Fig.~\ref{FIG1}.

For the photo- and electro-productions, respectively, we observe the strongly destructive and slightly constructive interferences between the $\Lambda$ and $\Sigma$ BKG contributions. The different interference pattern can be understood by additional phase factors in the electro-productions as shown in the diagram ($b$) in Fig.~\ref{FIG0} and the EM form factors in Fig.~\ref{FIG1}. On the contrary, we find that the WT contribution is relatively small and the KR one almost negligible for $g=1$. Thus, these $\Lambda$ and $\Sigma$ BKG contribution make almost all the strengths of the non-resonant BKG contributions. It also turns out that the EM form factors for the electro-production make the BKG strength much smaller by a factor $\sim10^{-2}$ and the BKG shape tilted to the lower $M_I$ region in comparison to the photo-production, due to the interference pattern changes by the EM form factors as mentioned. 

As shown in the line shapes in Fig.~\ref{FIG2}, we find that the lower-pole ($L$) contribution is almost unseen in comparison to the higher-pole ($H$) one for the photo- and electro-productions.  The reasons are as follows.   The dominant contributions to the present $K^*$ production are from the diagrams of ($a$) and ($c$) of Fig.~\ref{FIG0}. Their amplitudes contain the coupling combination
$|g_{K^*NL}g_{\pi\Sigma L}|$ for the lower and 
$|g_{K^*NH}g_{\pi\Sigma H}|$ for the higher pole contributions, respectively.  
Numerically, their ratio is $|g_{K^*NL}g_{\pi\Sigma L}|\approx0.9\times|g_{K^*NH}g_{\pi\Sigma H}|$~\cite{Khemchandani:2011mf}.  Therefore, the integrated strength of the $\pi^0\Sigma^0$ distribution is smaller for the lower than for the higher pole contributions. Moreover, the lower pole has much wider width as about four times larger than the one of the higher pole.  These explain the almost unseen effect from the lower pole in the $\pi^0\Sigma^0$ distribution.  In contrast, for the $K$ production, because of the relation 
$|g_{KNL}g_{\pi\Sigma L}|\approx1.6\times|g_{KNH}g_{\pi\Sigma H}|$~\cite{Khemchandani:2011mf}, the lower pole contributions are better seen in the $\pi^0 \Sigma^0$ distributions.

If the $K^*N\Lambda^*$ interaction is absent ($g=0$), the $K$ exchange dominates the reaction process and the $H$ peaks show small but finite strengths in comparison to the BKG contributions. When the $K^*N\Lambda^*$ interaction turns on with $g=1$, the situation changes drastically. As for the photo-production, the effects of the inclusion of the $K^*$ exchange gives about $0.1\,\mu$b increase in the $H$ peak position, due to a constructive interference between the $K$ and $K^*$ exchanges. The increase due to the $K^*$ exchange is more pronounced for the electro-production, since the $K^*$ EM form factors is larger than that of $K^*\to \gamma K$ transition one at $Q^2=2\,\mathrm{GeV}^2$ as shown in Fig.~\ref{FIG1}, resulting in the clearer peak signal at $M_I=1.43$ GeV over the BKG, as shown in the right panel of Fig.~\ref{FIG2}.

\begin{figure}[t]
\begin{tabular}{cc}
\includegraphics[width=6cm]{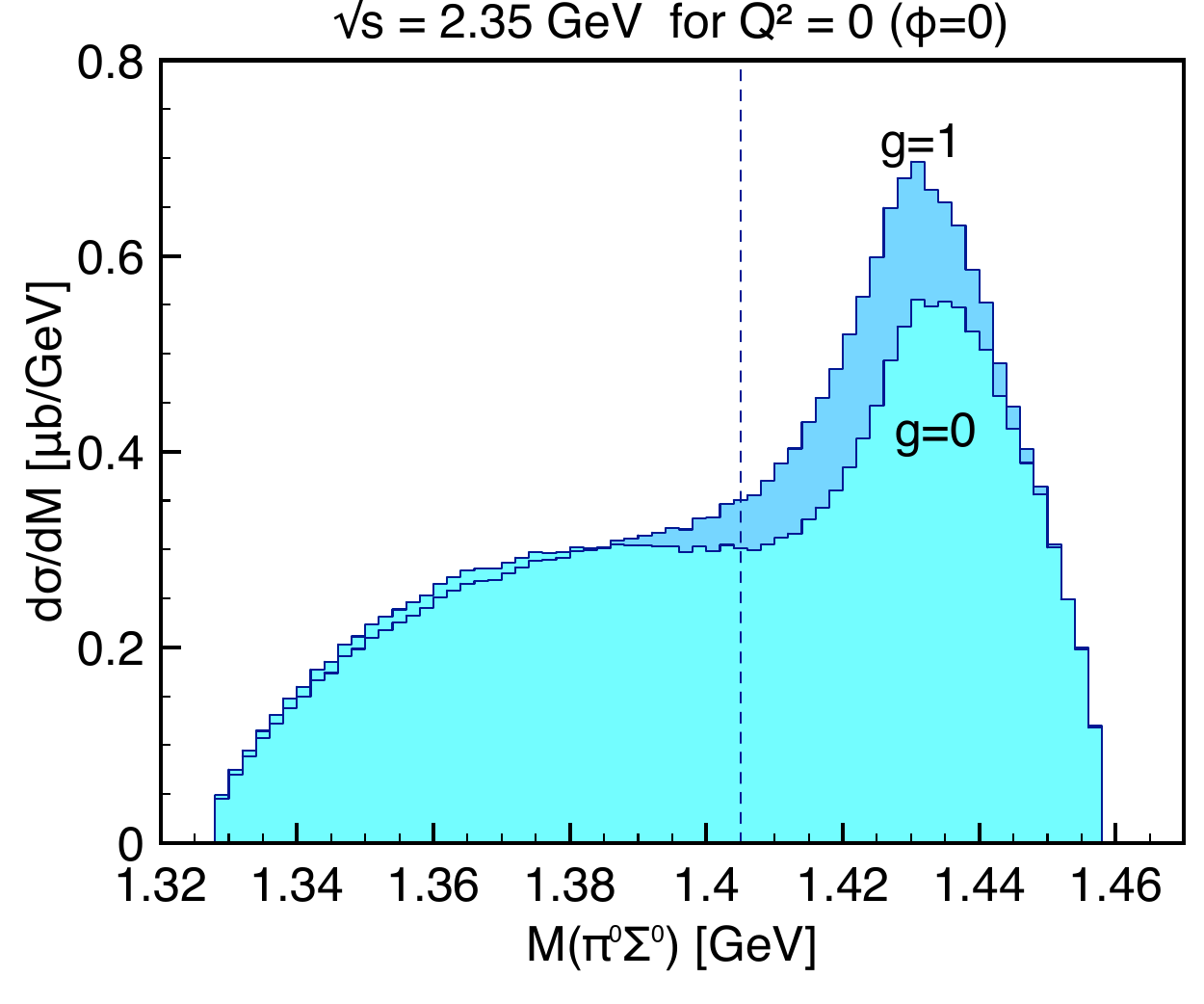}
\includegraphics[width=6cm]{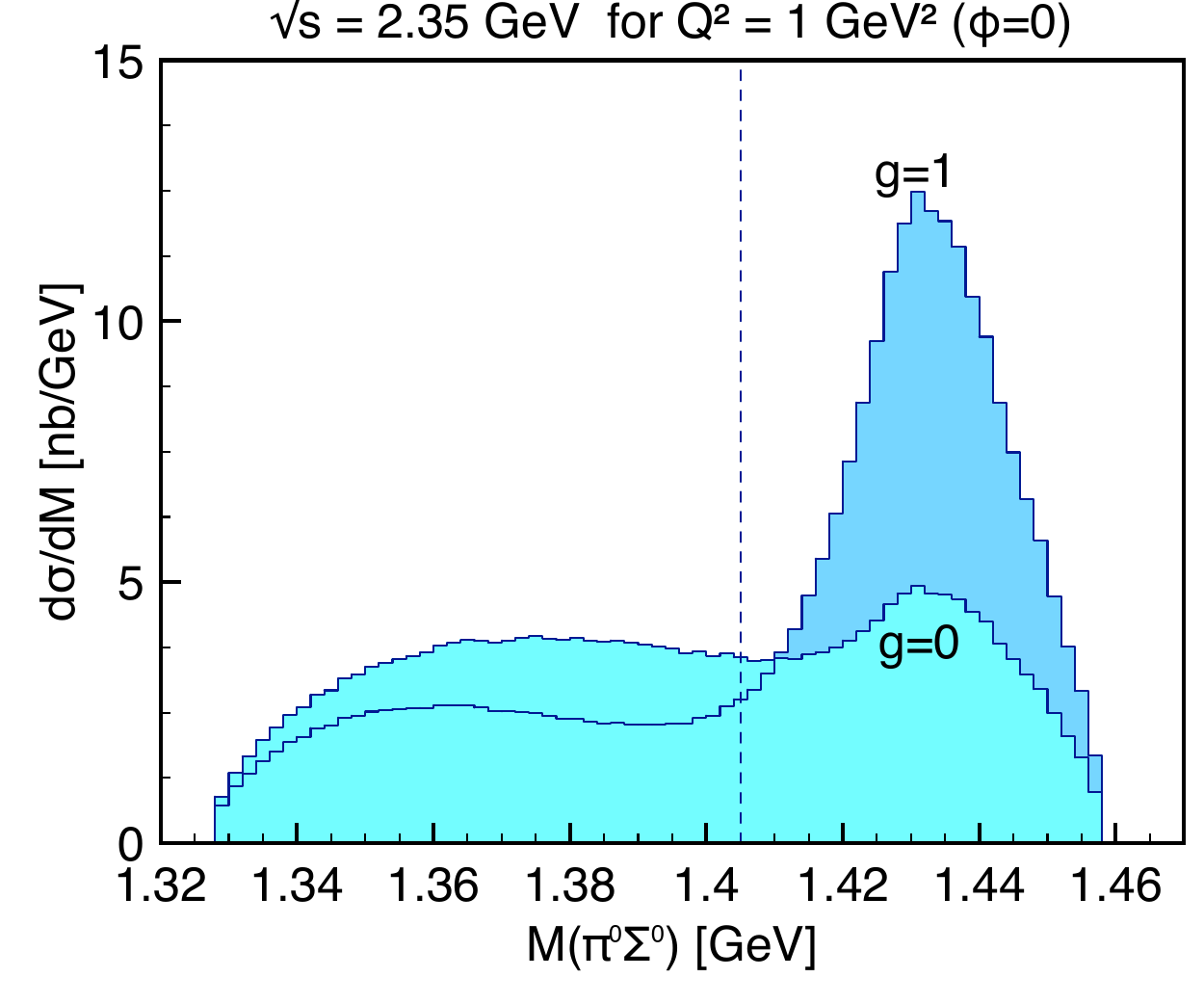}
\includegraphics[width=6cm]{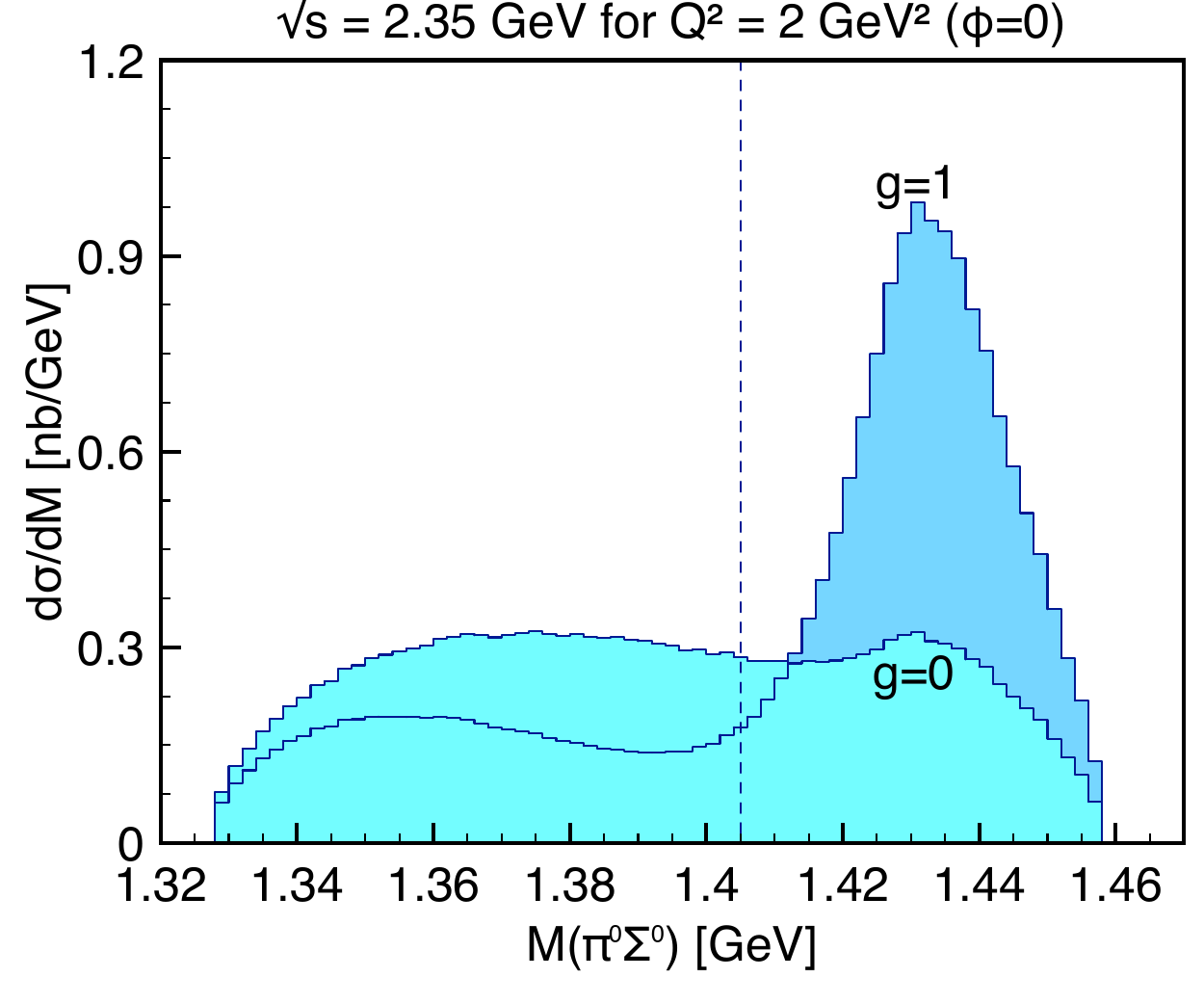}
\end{tabular}
\caption{(Color online) $\pi^0\Sigma^0$ invariant-mass plot $(d\sigma/dM_{\pi^0\Sigma^0})$ for $Q^2=(0,1.0,2.0)\,\mathrm{GeV}^2$ (left, middle, right column) at $\sqrt{s}=2.35$ GeV for $\phi=0$, using different choices of the parameter $g=1$ and $0$, which correspond to the cases with and without the $K^*$-exchange contribution in the $t$ channel. The vertical dashed line corresponds to the mass of $\Lambda(1405)$. See the text for the details.}   
\label{FIG3}
\begin{tabular}{cc}
\includegraphics[width=6cm]{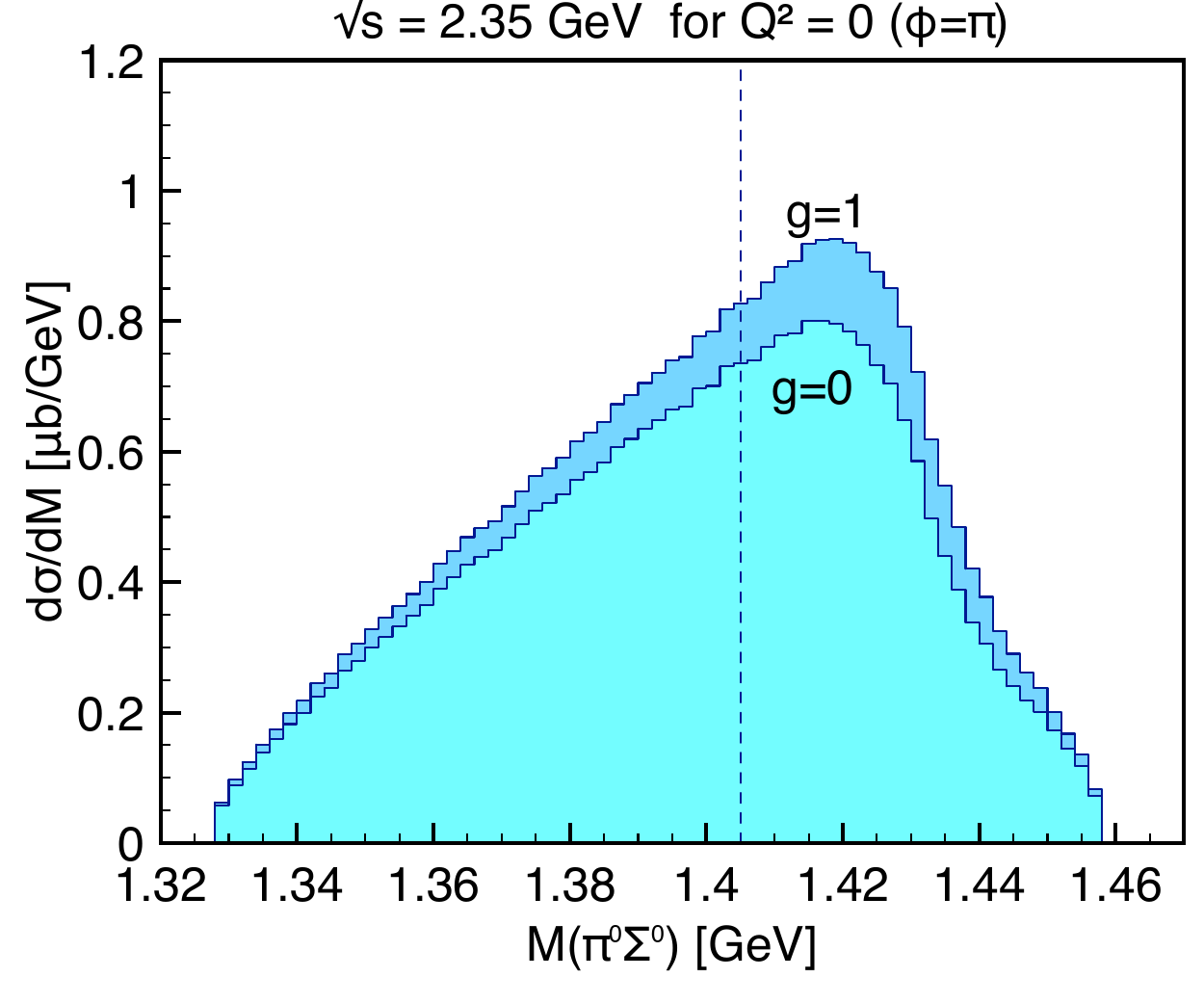}
\includegraphics[width=6cm]{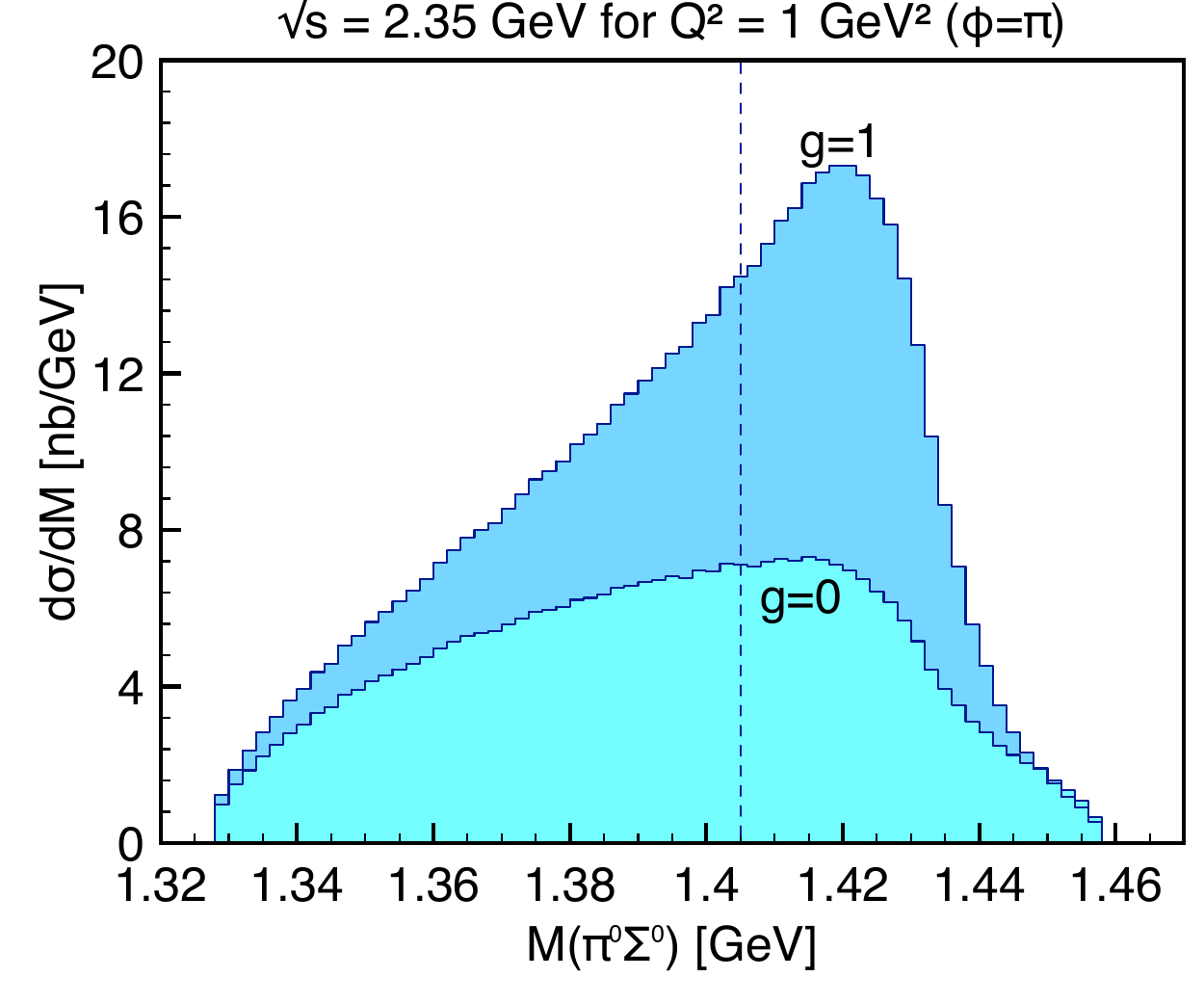}
\includegraphics[width=6cm]{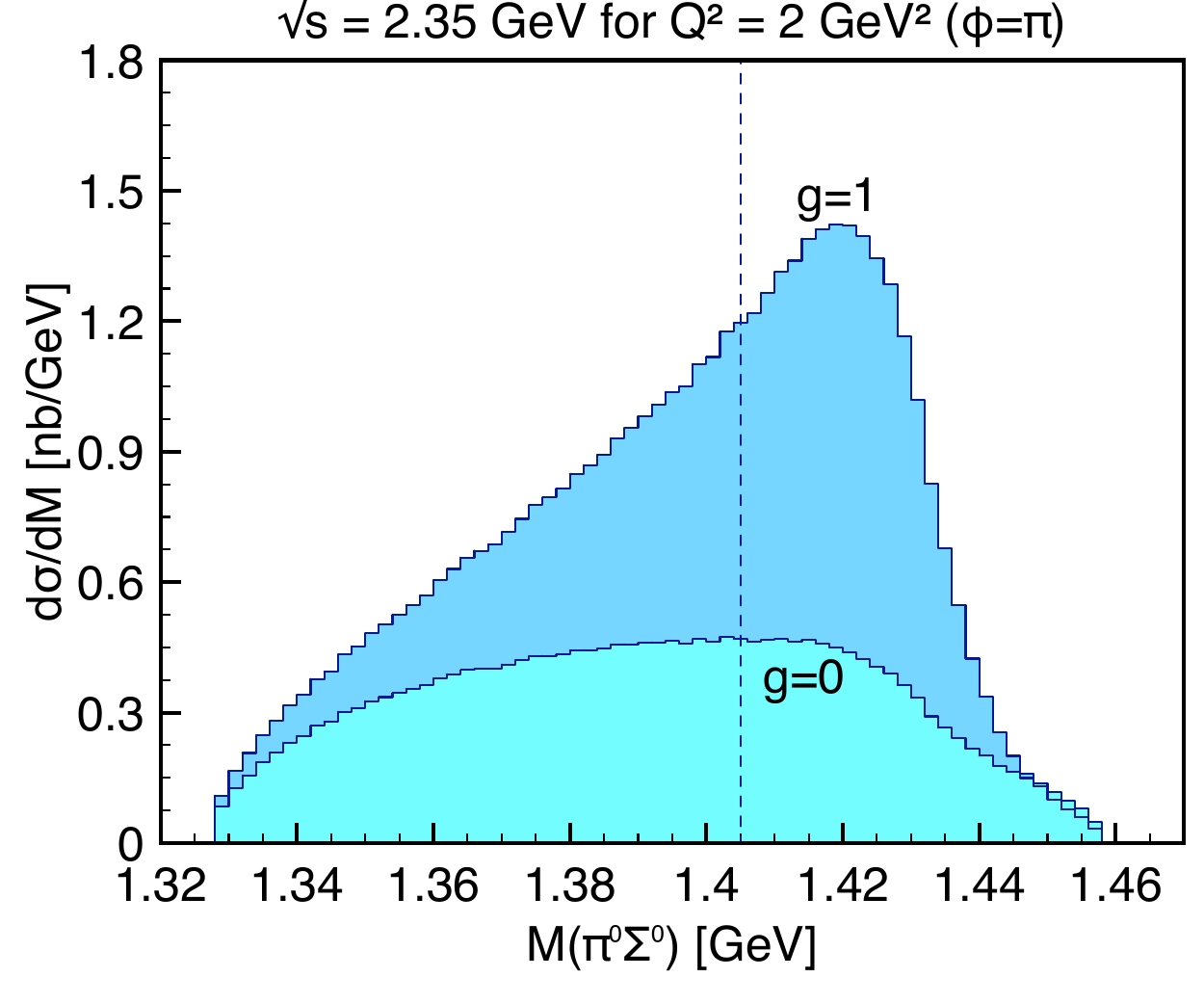}
\end{tabular}
\caption{(Color online) $\pi^0\Sigma^0$-invariant-mass plots in the same manner with Fig.~\ref{FIG3} for $\phi=\pi$. See the text for the details.}       
\label{FIG4}
\begin{tabular}{cc}
\includegraphics[width=6cm]{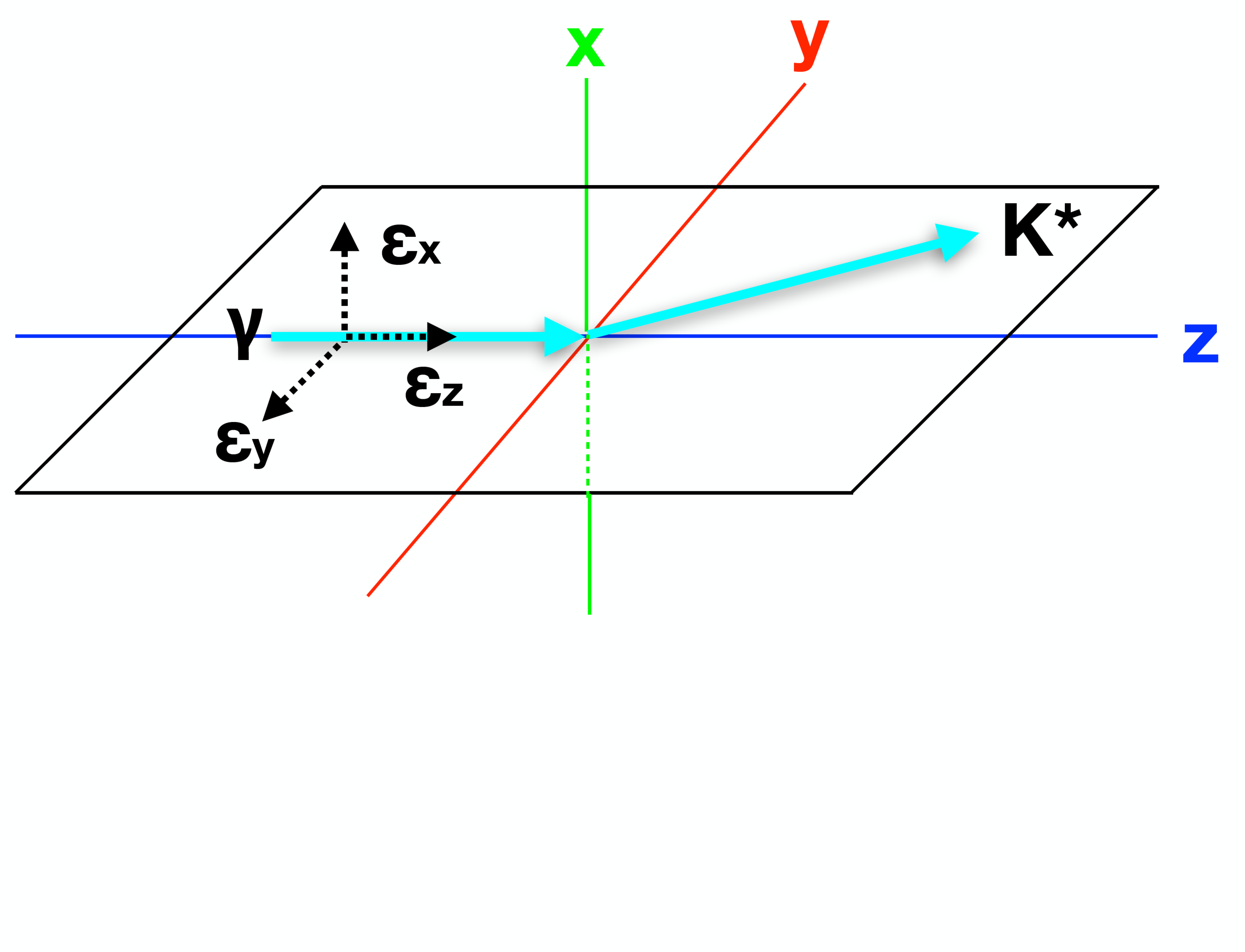}
\includegraphics[width=6cm]{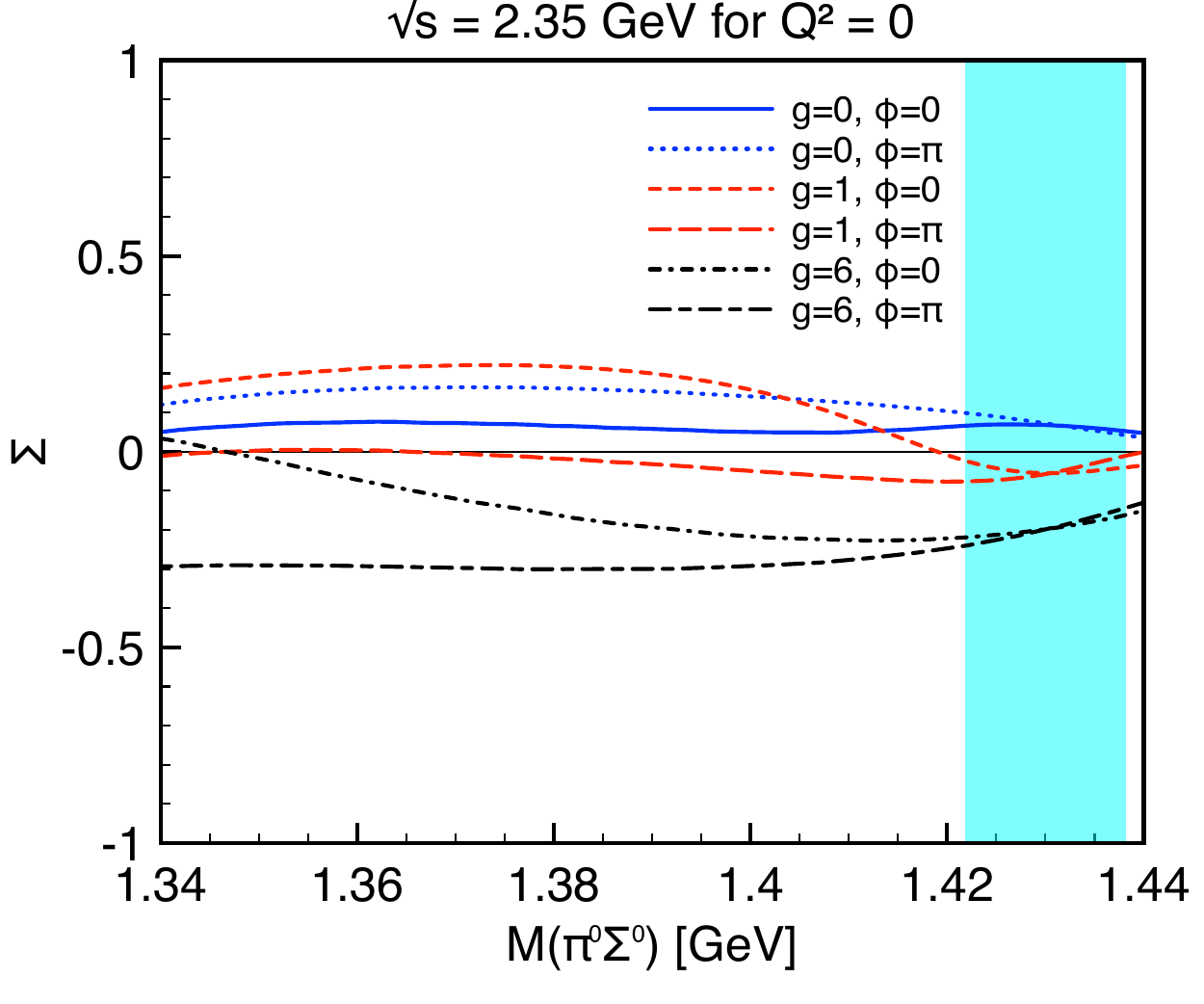}
\includegraphics[width=6cm]{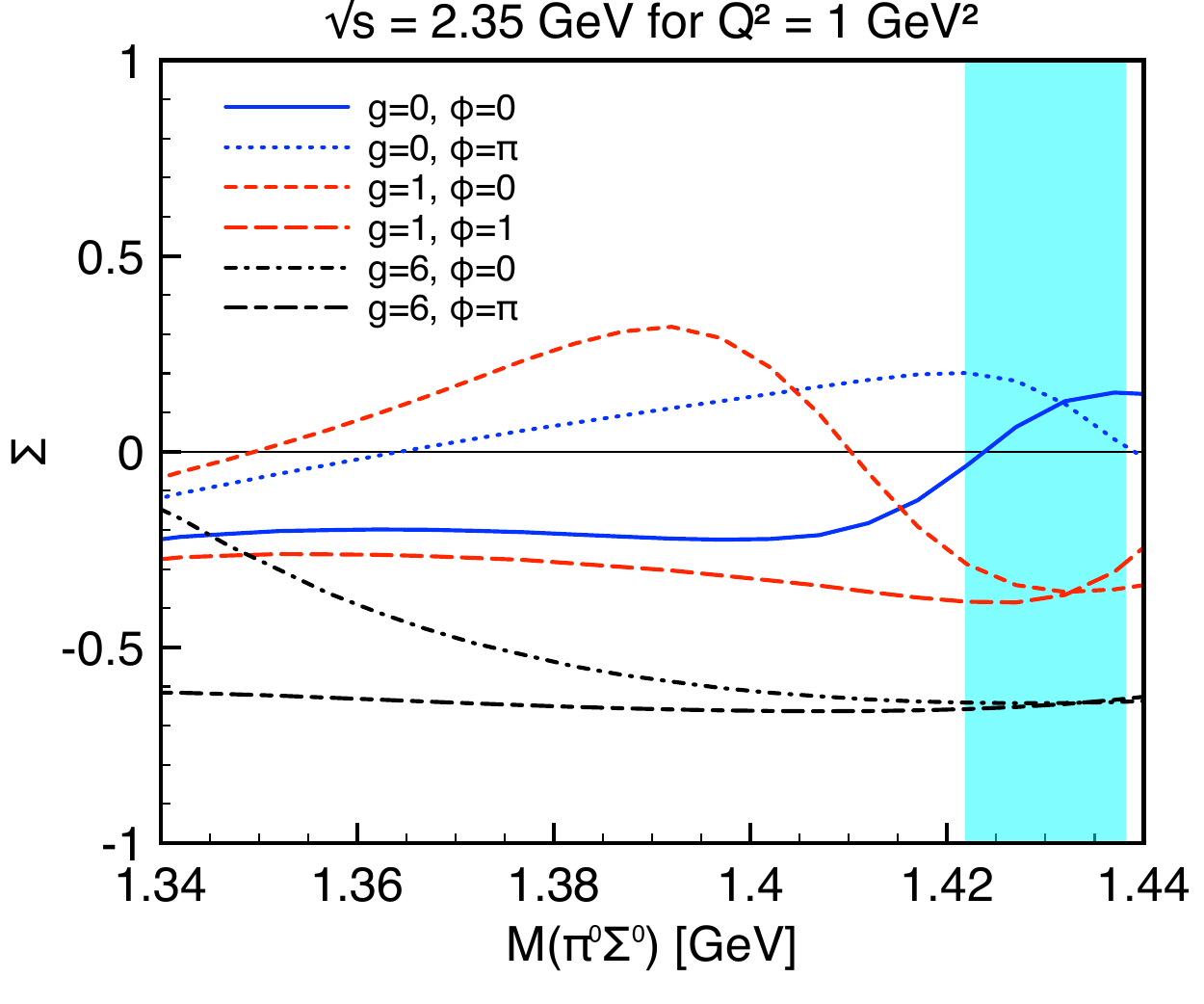}
\end{tabular}
\caption{(Color online) Left: Definition of the photon polarizations and the relevant momenta for $\Sigma$, defined in Eq.~(\ref{eq:SIG}). Middle: Numerical results for $\Sigma$ with $Q^2=0$ varying $g$ and $\phi$. The vertical shaded area represents the higher-pole peak region. Right: The same with $Q^2=1\,\mathrm{GeV}^2$.}       
\label{FIG56}
\end{figure}

Now, we present the numerical results for the invariant-mass line shape including all the contributions, varying the phase angle $\phi$ defined in Eq.~(\ref{eq:TAMP}) as a free parameter. For simplicity, we only consider two cases with $\phi=0$ and $\pi$, and those results are shown in Fig.~\ref{FIG3} and Fig.~\ref{FIG4}, respectively. The vertical dashed line indicates $M_I=1.405$ GeV. As for the case with $\phi=0$ and $Q^2=0$ (photo-production) shown in the left panel of Fig.~\ref{FIG3}, it turns out that the $H$ and BKG contributions interfere constructively, showing a very pronounced peak in the vicinity of $M_I=1.43$ GeV, and a broad bump for $M_I\lesssim1.4$ GeV, due to the BKG contributions. Again, we observe that the inclusion of the $K^*$ exchange gives small increases in the peak region, but the line shape remains almost the same.  As for the electro-production for $Q^2=(1,2)\,\mathrm{GeV}^2$ in the (middle, right) panels of Fig.~\ref{FIG3}, the $H$ peak becomes enhanced considerably, according to the larger $K^*\to\gamma K$ transition form factor than the $K^*$ EM one as explained previously. From these numerical results, we conclude that the $\Lambda^*$ peak can be clearly observed both for the photo- and electro-productions, if the $K^*N\Lambda^*$ interaction is finite. On the contrary, if the interaction is negligible, the clear peak signal survives only for the photo-production. Hence, the electro-production is better to test the effects of the $K^*N\Lambda^*$ interaction. In Fig.~\ref{FIG4}, we also draw the same curves for $\phi=\pi$ and observe similar tendencies, although the $H$ and the BKG contributions interfere destructively. It turns out that this destructive interference makes the $H$ peak more dubious for $g=0$ than that with $\phi=0$. In general, the peak signal decrease with respect to $Q^2$.

Finally, we are in a position to define a polarization observable as a function of $M_I$, which is similar to the photon-beam asymmetry, and responsible for estimating the effects of the $K^*$-exchange contribution:
\begin{equation}
\label{eq:SIG}
\Sigma(M_I)=\frac{d\sigma_x/dM_I-d\sigma_y/dM_I-d\sigma_z/dM_I}
{d\sigma_x/dM_I+d\sigma_y/dM_I+d\sigma_z/dM_I},
\end{equation}
where the subscripts $x$, $y$, and $z$ denote the photon linear polarizations $\vec{\varepsilon}_x$ (perpendicular), $\vec{\varepsilon}_y$ (parallel), and $\vec{\varepsilon}_z$ (longitudinal). Here, the incident photon and outgoing $K^*$ determines the $y$-$z$ reaction plane in the center-of-mass frame, as shown in the left panel of Fig.~\ref{FIG56}. Note that the third term $d\sigma_z/dM_I$ in Eq.~(\ref{eq:SIG}) only exists for the electro-production. In the middle panel of Fig.~\ref{FIG56}, we depict the numerical results of $\Sigma$ for the photo-production $(Q^2=0)$, using various choices of the model parameters. The shaded area represents the $H$ peak region. 

As for $g=0$, i.e., the reaction process is dominated by the $K$-exchange and the values of $\Sigma$ are \textit{positive}. This observation indicates that the $K$-exchange contribution almost disappears for the parallel polarization $d\sigma_y/dM_I\sim0$, due to the antisymmetric-tensor structure of $\mathcal{M}_d$ in Eq.~(\ref{eq:AMP1}) $\propto\vec{\epsilon}_\gamma\times\vec{k}_{K^*}$. In contrast, if we take into account the $K^*$-exchange contribution with $g=1$, the perpendicular contribution $d\sigma_x/dM_I$ becomes negligible, resulting in the negative $\Sigma$ values in the peak region as shown in the left panel of Fig.~(\ref{FIG56}). Again, this behavior can be explained by the Lorentz structure of $\mathcal{M}_{a,b,c}$ in Eq.~(\ref{eq:AMP1}) $\propto\vec{\epsilon}_\gamma\cdot\vec{k}_{K^*}$. As shown by the curves for $g=6$, this tendency becomes more apparent, since the higher-pole contribution dominates the cross section for the larger $g$ values.

In the right panel of Fig.~\ref{FIG56}, we show the numerical results of $\Sigma$ for the electro-production in the same manner with the photo-production. Since the spin-$0$ scalar component of the longitudinal photon polarization enhances the spin-$1$ exchange contribution, i.e., the $K^*$-exchange one, the third term in Eq.~(\ref{eq:SIG}) is magnified. Hence, as shown in the right panel of Fig.~\ref{FIG56}, the signals in the peak regions become more obvious for $g=1$. Note that we observe the same tendency with respect to the larger $g$ for the electro-production as well.

Taking these numerical results of $\Sigma$ into account, if one observes the positive $\Sigma$ values for the photo- and electro-productions in the vicinity of the peak region in experiments, it indicates the $K^*$-exchange contribution must be negligible. In contrast, one can conclude that $K^*N\Lambda^*$ interaction is sizable, when the negative values of $\Sigma$ are measured. Consequently, by examining the sign of $\Sigma$ in the peak region experimentally, one can estimate the strength of $g_{K^*N\Lambda^*}$ uniquely. 

\section{Summary}
In the present work, we investigated the $K^*$ photo- and electro-productions via $\gamma^{*}p\to K^{*+}\pi^0\Sigma^0$, in which $\Lambda(1405)\equiv\Lambda^*$ appears as a dominant hyperon resonance near the threshold. Moreover, this reaction process does not contain $\Sigma^*(1385)$, which can interfere with $\Lambda^*$ in the charged channels such as $\gamma^{*}p\to K^{*+}\pi^\mp\Sigma^\pm$, resulting in a clear signal only from $\Lambda^*$. The effective Lagrangian method was employed at the tree-level Born approximation. We used the phenomenological strong and electromagnetic (EM) form factors for the relevant hadrons, and the EM form factors for $\Lambda^*$ was parameterized by using the information from the ChUM results. The $K^*\to \gamma K$ transition form factor was devised from the kaon light-front wave function, which was computed by the nonlocal chiral-quark model in our previous works, whereas the proton and vector kaon EM form factors were taken from available data.  Focusing on the two-pole structure scenario of $\Lambda^*$ and the rarely-known $K^*N\Lambda^*$ coupling constant $g_{K^*N\Lambda^*}$, we provide the numerical results for the invariant-mass line shape and the photon linear-polarization observable $\Sigma$. We list important observations in the present work as follows:
\begin{itemize}
\item The lower-pole peak ($L$) turns out to be unseen, due to its larger width $\Gamma_L\approx126$ MeV $\gg$ $\Gamma_H\approx30$ MeV, when we resort to the ChUM results as theory inputs. It also turns out that the background (BKG) contribution comes mainly from the destructive and constructive interferences between $\Lambda$ and $\Sigma$ ground states for the photo- and electro-production, respectively. This different interference pattern can be understood by their EM  form factors, which modify the phases between the invariant amplitudes.  
\item We  find that the $\Lambda^*$ peak is clearly observed for the photo- and electro-productions with the finite $K^*N\Lambda^*$ interaction, whereas the peak survives only for the photo-production when we ignore the interaction. These interesting behaviors of the peak can be understood by the different $Q^2$ dependences in the $K^*$ electromagnetic and $K^*\to\gamma K$ transition form factors. Taking into account these observations, if the line shapes of the photo- and electro-production exhibit considerable differences in the peak region, it can be said that the $K^*NH$ interaction is not important. 
\item In order to estimate the strength of the $K^*N\Lambda^*$ interaction more precisely, we suggest a photon-polarization observable $\Sigma$.  As for the photo-production, where the longitudinal polarization does not exists, the $K$- and $K^*$-exchange contributions are almost canceled out for the parallel and perpendicular polarizations, respectively. Hence, by construction, $\Sigma$ becomes positive and negative in the peak region $M_I\sim1.43$ GeV for the finite and negligible $K^*NH$ interaction, respectively. When the longitudinal photon polarization, which gives the spin-$0$ scalar (natural spin-parity) component, comes into play for the electro-production, it enhances the $K^*$-exchange contribution and the negative signal becomes more obvious. Thus, by examining the polarization observable $\Sigma$ in the vicinity of the $H$ peak, one can estimate the strength of $g_{K^*NH}$ in the experiments. 
\end{itemize}

Although there are several theoretical uncertainties, such as the parameterization of the relevant form factors, we have provided theoretical results for understanding the nature of $\Lambda^*$ produced with the vector kaon and also suggested how to estimate the $K^*N\Lambda^*$ interaction strength uniquely for the future experiments by the LEPS and CLAS collaborations. More realistic production process including the decay of $K^*\to\pi K$ in the four-body phase space and the $\Lambda^*$ production with the pseudoscalar-meson beam, such as $\pi N\to K\pi\Sigma$, are being studied, and the related works will appear elsewhere. 
\section*{Acknowledgment}
The authors thank K.~P.~Khemchandani, A.~Martinez Torres, and J.~K.~Ahn for fruitful discussions. The work of S.i.N. was supported in part by the National Research Foundation of Korea (NRF) grants (No.~2018R1A5A1025563 and No.~2019R1A2C1005697). The work of A.H. is supported in part by the Grants-in-Aid for Scientific Research (No.~JP17K05441 (C)).


\begin{thebibliography}{99}
\bibitem{Leutwyler:1993iq} 
  H.~Leutwyler,
  Annals Phys.\  {\bf 235}, 165 (1994).
\bibitem{Jido:2003cb} 
  D.~Jido, J.~A.~Oller, E.~Oset, A.~Ramos and U.~G.~Meissner,
  Nucl.\ Phys.\ A {\bf 725}, 181 (2003).
\bibitem{Magas:2005vu} 
  V.~K.~Magas, E.~Oset and A.~Ramos,
  Phys.\ Rev.\ Lett.\  {\bf 95}, 052301 (2005).
\bibitem{Jido:2002zk} 
  D.~Jido, E.~Oset and A.~Ramos,
  Phys.\ Rev.\ C {\bf 66}, 055203 (2002).
\bibitem{Hyodo:2011ur} 
  T.~Hyodo and D.~Jido,
  Prog.\ Part.\ Nucl.\ Phys.\  {\bf 67}, 55 (2012).
\bibitem{Nam:2017yeg} 
  S.~i.~Nam,
  Phys.\ Rev.\ D {\bf 96}, no. 7, 076021 (2017).
\bibitem{Frazer:1964zz} 
  W.~R.~Frazer and A.~W.~Hendry,
  Phys.\ Rev.\  {\bf 134}, B1307 (1964).
\bibitem{Roca:2013av} 
  L.~Roca and E.~Oset,
  Phys.\ Rev.\ C {\bf 87}, no. 5, 055201 (2013).
\bibitem{Roca:2013cca} 
  L.~Roca and E.~Oset,
  Phys.\ Rev.\ C {\bf 88}, no. 5, 055206 (2013).
\bibitem{Hall:2014uca} 
  J.~M.~M.~Hall \textit{et al.},
  Phys.\ Rev.\ Lett.\  {\bf 114}, no. 13, 132002 (2015).
\bibitem{Cieply:2016jby} 
  A.~Cieplý, M.~Mai, U.~G.~Meißner and J.~Smejkal,
  Nucl.\ Phys.\ A {\bf 954}, 17 (2016).
\bibitem{Lu:2013nza} 
  H.~Y.~Lu {\it et al.} [CLAS Collaboration],
  Phys.\ Rev.\ C {\bf 88}, 045202 (2013).
\bibitem{Niiyama:2009zza} 
  M.~Niiyama [LEPS TPC Collaboration],
  Nucl.\ Phys.\ A {\bf 827}, 261C (2009).
\bibitem{Moriya:2013hwg} 
  K.~Moriya {\it et al.} [CLAS Collaboration],
  Phys.\ Rev.\ C {\bf 88}, 045201 (2013)
  [Addendum-ibid.\ C {\bf 88}, 049902 (2013)].
\bibitem{Moriya:2013eb} 
  K.~Moriya {\it et al.} [CLAS Collaboration],
  Phys.\ Rev.\ C {\bf 87}, no. 3, 035206 (2013).
\bibitem{Sekihara:2008qk} 
  T.~Sekihara, T.~Hyodo and D.~Jido,
  Phys.\ Lett.\ B {\bf 669}, 133 (2008).
\bibitem{Khemchandani:2011mf} 
  K.~P.~Khemchandani, A.~Martinez Torres, H.~Kaneko, H.~Nagahiro and A.~Hosaka,
  Phys.\ Rev.\ D {\bf 84}, 094018 (2011).
\bibitem{Nam:2013nfa} 
  S.~i.~Nam,
  J.\ Phys.\ G {\bf 40}, 115001 (2013).
\bibitem{Kaskulov:2003bg} 
  M.~M.~Kaskulov and P.~Grabmayr,
  Eur.\ Phys.\ J.\ A {\bf 19}, 157 (2004).
\bibitem{Hawes:1998bz} 
  F.~T.~Hawes and M.~A.~Pichowsky,
  Phys.\ Rev.\ C {\bf 59}, 1743 (1999).
\bibitem{Khodjamirian:1997tk} 
  A.~Khodjamirian,
  Eur.\ Phys.\ J.\ C {\bf 6}, 477 (1999).
\bibitem{Nam:2006au} 
  S.~i.~Nam, H.~C.~Kim, A.~Hosaka and M.~M.~Musakhanov,
  Phys.\ Rev.\ D {\bf 74}, 014019 (2006).
\bibitem{Nam:2006sx} 
  S.~i.~Nam and H.~C.~Kim,
  Phys.\ Rev.\ D {\bf 74}, 076005 (2006).
\bibitem{Keller:2011nt} 
  D.~Keller {\it et al.} [CLAS Collaboration],
  Phys.\ Rev.\ D {\bf 83}, 072004 (2011).
\bibitem{Dhir:2009ax} 
  R.~Dhir and R.~C.~Verma,
  Eur.\ Phys.\ J.\ A {\bf 42}, 243 (2009).
\bibitem{Wang:2017tpe} 
  A.~C.~Wang, W.~L.~Wang, F.~Huang, H.~Haberzettl and K.~Nakayama,
  Phys.\ Rev.\ C {\bf 96}, 035206 (2017).
\bibitem{Huang:2018qym} 
  F.~Huang, A.~C.~Wang, W.~L.~Wang, H.~Haberzettl and K.~Nakayama,
  Few Body Syst.\  {\bf 59}, no. 5, 84 (2018).
\bibitem{Ryu:2016jmv} 
  S.~Y.~Ryu {\it et al.} [LEPS Collaboration],
  Phys.\ Rev.\ Lett.\  {\bf 116}, no. 23, 232001 (2016).
\bibitem{Olive:2016xmw} 
  C.~Patrignani {\it et al.} [Particle Data Group],
  Chin.\ Phys.\ C {\bf 40}, 100001 (2016).
\bibitem{Stoks:1999bz} 
  V.~G.~J.~Stoks and T.~A.~Rijken,
  Phys.\ Rev.\ C {\bf 59}, 3009 (1999).
\bibitem{Davidson:2001rk}
  R.~M.~Davidson and R.~Workman,
  Phys.\ Rev.\  C {\bf 63}, 025210 (2001).
\bibitem{Perdrisat:2006hj} 
  C.~F.~Perdrisat, V.~Punjabi and M.~Vanderhaeghen,
  Prog.\ Part.\ Nucl.\ Phys.\  {\bf 59}, 694 (2007).
\bibitem{Berger:2004yi} 
  K.~Berger, R.~F.~Wagenbrunn and W.~Plessas,
  Phys.\ Rev.\ D {\bf 70}, 094027 (2004).
\bibitem{Nam:2003uf} 
  S.~i.~Nam, A.~Hosaka and H.~C.~Kim,
  Phys.\ Lett.\ B {\bf 579}, 43 (2004).
\bibitem{Nam:2003ch} 
  S.~i.~Nam, H.~C.~Kim, T.~Hyodo, D.~Jido and A.~Hosaka,
  J.\ Korean Phys.\ Soc.\  {\bf 45}, 1466 (2004).
\end{thebibliography}
\end{document}